
%
\newbox\leftpage \newdimen\fullhsize \newdimen\hstitle \newdimen\hsbody
\tolerance=1000\hfuzz=2pt
\def\printertype{ps: }
\def\qms{\def\printertype{qms: }
\ifx\answ\bigans\else\voffset=-.4truein\hoffset=.125truein\fi}
\def\bigans{b }
%
\def\answ{b }
%
\font\titlerm=cmr10 scaled\magstep3 \font\titlerms=cmr7 scaled\magstep3
\font\titlermss=cmr5 scaled\magstep3 \font\titlei=cmmi10 scaled\magstep3
\font\titleis=cmmi7 scaled\magstep3 \font\titleiss=cmmi5 scaled\magstep3
\font\titlesy=cmsy10 scaled\magstep3 \font\titlesys=cmsy7 scaled\magstep3
\font\titlesyss=cmsy5 scaled\magstep3 \font\titleit=cmti10 scaled\magstep3
\skewchar\titlei='177 \skewchar\titleis='177 \skewchar\titleiss='177
\skewchar\titlesy='60 \skewchar\titlesys='60 \skewchar\titlesyss='60
\def\titlefont{\def\rm{\fam0\titlerm}
\textfont0=\titlerm \scriptfont0=\titlerms \scriptscriptfont0=\titlermss
\textfont1=\titlei \scriptfont1=\titleis \scriptscriptfont1=\titleiss
\textfont2=\titlesy \scriptfont2=\titlesys \scriptscriptfont2=\titlesyss
\textfont\itfam=\titleit \def\it{\fam\itfam\titleit} \rm}
\font\absrm=cmr10 scaled\magstep1 \font\absrms=cmr7 scaled\magstep1
\font\absrmss=cmr5 scaled\magstep1 \font\absi=cmmi10 scaled\magstep1
\font\absis=cmmi7 scaled\magstep1 \font\absiss=cmmi5 scaled\magstep1
\font\abssy=cmsy10 scaled\magstep1 \font\abssys=cmsy7 scaled\magstep1
\font\abssyss=cmsy5 scaled\magstep1 \font\absbf=cmbx10 scaled\magstep1
\skewchar\absi='177 \skewchar\absis='177 \skewchar\absiss='177
\skewchar\abssy='60 \skewchar\abssys='60 \skewchar\abssyss='60
\font\bigit=cmti10 scaled \magstep1
\def\abstractfont{\def\rm{\fam0\absrm}
\textfont0=\absrm \scriptfont0=\absrms \scriptscriptfont0=\absrmss
\textfont1=\absi \scriptfont1=\absis \scriptscriptfont1=\absiss
\textfont2=\abssy \scriptfont2=\abssys \scriptscriptfont2=\abssyss
\textfont\itfam=\bigit \def\it{\fam\itfam\bigit}
\textfont\bffam=\absbf \def\bf{\fam\bffam\absbf} \rm} 
\def\tenpoint{\def\rm{\fam0\tenrm}
\textfont0=\tenrm \scriptfont0=\sevenrm \scriptscriptfont0=\fiverm
\textfont1=\teni  \scriptfont1=\seveni  \scriptscriptfont1=\fivei
\textfont2=\tensy \scriptfont2=\sevensy \scriptscriptfont2=\fivesy
\textfont\itfam=\tenit \def\it{\fam\itfam\tenit}
\textfont\bffam=\tenbf \def\bf{\fam\bffam\tenbf} \rm}
%
%
\abstractfont
\baselineskip=20pt plus 2pt minus 1pt
\hsbody=\hsize \hstitle=\hsize 
%
%
\catcode`\@=11 
\newcount\yearltd\yearltd=\year\advance\yearltd by -1900

\def\Title#1#2#3#4{\nopagenumbers\abstractfont
\hsize=\hstitle\rightline{#1}
\hsize=\hstitle\rightline{#2}
\vskip 1in\centerline{\titlefont #3}
\centerline{\titlefont #4}\abstractfont\vskip .5in\pageno=0}
\def\Date#1{{\vfill\leftline{#1}\tenpoint\supereject\global\hsize=\hsbody%
\footline={\hss\tenrm\folio\hss}}}
%

\def\draftmode{\message{ DRAFTMODE }\def\draftdate{{\rm preliminary draft:
\number\month/\number\day/\number\yearltd\ \ \hourmin}}%
\headline={\hfil\draftdate}\writelabels\baselineskip=20pt plus 2pt minus 2pt
 {\count255=\time\divide\count255 by 60 \xdef\hourmin{\number\count255}
  \multiply\count255 by-60\advance\count255 by\time
  \xdef\hourmin{\hourmin:\ifnum\count255<10 0\fi\the\count255}}}
\def\nolabels{\def\wrlabel##1{}\def\eqlabel##1{}\def\reflabel##1{}}
\def\writelabels{\def\wrlabel##1{\leavevmode\vadjust
{\rlap{\smash{\line{{\escapechar=` \hfill\rlap{\hskip.03in\string##1}}}}}}}%
\def\eqlabel##1{{\escapechar=` \rlap{\hskip.09in\string##1}}}%
\def\reflabel##1{\noexpand\llap{\string\string\string##1}}}
\nolabels
%
\global\newcount\secno \global\secno=0
\global\newcount\meqno \global\meqno=1
\def\newsec#1{\global\advance\secno by1\message{(\the\secno. #1)}
\global\subsecno=0\xdef\secsym{\the\secno.}\global\meqno=1
\bigbreak\bigskip\noindent{\bf\the\secno. #1}\writetoca{{\secsym} {#1}}
\par\nobreak\medskip\nobreak}
\xdef\secsym{}
\global\newcount\subsecno \global\subsecno=0
\def\subsec#1{\global\advance\subsecno by1\message{(\secsym\the\subsecno. #1)}
\bigbreak\noindent{\it\secsym\the\subsecno. #1}\writetoca{\string\quad
{\secsym\the\subsecno.} {#1}}\par\nobreak\medskip\nobreak}
\def\appendix#1#2{\global\meqno=1\global\subsecno=0\xdef\secsym{\hbox{#1.}}
\bigbreak\bigskip\noindent{\bf Appendix #1. #2}\message{(#1. #2)}
\writetoca{Appendix {#1.} {#2}}\par\nobreak\medskip\nobreak}
%
%
\def\eqnn#1{\xdef #1{(\secsym\the\meqno)}\writedef{#1\leftbracket#1}%
\global\advance\meqno by1\wrlabel#1}
\def\eqna#1{\xdef #1##1{\hbox{$(\secsym\the\meqno##1)$}}
\writedef{#1\numbersign1\leftbracket#1{\numbersign1}}%
\global\advance\meqno by1\wrlabel{#1$\{\}$}}
\def\eqn#1#2{\xdef #1{(\secsym\the\meqno)}\writedef{#1\leftbracket#1}%
\global\advance\meqno by1$$#2\eqno#1\eqlabel#1$$}
%
\newskip\footskip\footskip14pt plus 1pt minus 1pt 
\def\f@@t{\baselineskip\footskip\bgroup\aftergroup\@foot\let\next}
\setbox\strutbox=\hbox{\vrule height9.5pt depth4.5pt width0pt}
\global\newcount\ftno \global\ftno=0
\def\foot{\global\advance\ftno by1\footnote{$^{\the\ftno}$}}
%
\newwrite\ftfile
\def\footend{\def\foot{\global\advance\ftno by1\chardef\wfile=\ftfile
$^{\the\ftno}$\ifnum\ftno=1\immediate\openout\ftfile=foots.tmp\fi%
\immediate\write\ftfile{\noexpand\smallskip%
\noexpand\item{f\the\ftno:\ }\pctsign}\findarg}%
\def\footatend{\vfill\eject\immediate\closeout\ftfile{\parindent=20pt
\centerline{\bf Footnotes}\nobreak\bigskip\input foots.tmp }}}
\def\footatend{}
%
%
\global\newcount\refno \global\refno=1
\newwrite\rfile
\def\ref{[\the\refno]\nref}
\def\nref#1{\xdef#1{[\the\refno]}\writedef{#1\leftbracket#1}%
\ifnum\refno=1\immediate\openout\rfile=refs.tmp\fi
\global\advance\refno by1\chardef\wfile=\rfile\immediate
\write\rfile{\noexpand\item{#1\ }\reflabel{#1\hskip.31in}\pctsign}\findarg}
\def\findarg#1#{\begingroup\obeylines\newlinechar=`\^^M\pass@rg}
{\obeylines\gdef\pass@rg#1{\writ@line\relax #1^^M\hbox{}^^M}%
\gdef\writ@line#1^^M{\expandafter\toks0\expandafter{\striprel@x #1}%
\edef\next{\the\toks0}\ifx\next\em@rk\let\next=\endgroup\else\ifx\next\empty%
\else\immediate\write\wfile{\the\toks0}\fi\let\next=\writ@line\fi\next\relax}}
\def\striprel@x#1{} \def\em@rk{\hbox{}}

\def\addref#1{\immediate\write\rfile{\noexpand\item{}#1}} 
\def\startrefs#1{\immediate\openout\rfile=refs.tmp\refno=#1}
\def\xref{\expandafter\xr@f}\def\xr@f[#1]{#1}
\def\refs#1{[\r@fs #1{\hbox{}}]}
\def\r@fs#1{\edef\next{#1}\ifx\next\em@rk\def\next{}\else
\ifx\next#1\xref #1\else#1\fi\let\next=\r@fs\fi\next}
%

%
\newwrite\ffile\global\newcount\figno \global\figno=1
\def\fig{fig.~\the\figno\nfig}
\def\nfig#1{\xdef#1{fig.~\the\figno}%
\writedef{#1\leftbracket fig.\noexpand~\the\figno}%
\ifnum\figno=1\immediate\openout\ffile=figs.tmp\fi\chardef\wfile=\ffile%
\immediate\write\ffile{\noexpand\medskip\noexpand\item{Fig.\ \the\figno. }
\reflabel{#1\hskip.55in}\pctsign}\global\advance\figno by1\findarg}
\def\xfig{\expandafter\xf@g}\def\xf@g fig.\penalty\@M\ {}
\def\figs#1{figs.~\f@gs #1{\hbox{}}}
\def\f@gs#1{\edef\next{#1}\ifx\next\em@rk\def\next{}\else
\ifx\next#1\xfig #1\else#1\fi\let\next=\f@gs\fi\next}
\newwrite\lfile
{\escapechar-1\xdef\pctsign{\string\%}\xdef\leftbracket{\string\{}
\xdef\rightbracket{\string\}}\xdef\numbersign{\string\#}}

\def\writestop{\def\writestoppt{\immediate\write\lfile{\string\pageno%
\the\pageno\string\startrefs\leftbracket\the\refno\rightbracket%
\string\def\string\secsym\leftbracket\secsym\rightbracket%
\string\secno\the\secno\string\meqno\the\meqno}\immediate\closeout\lfile}}
\def\writestoppt{}\def\writedef#1{}
\def\seclab#1{\xdef #1{\the\secno}\writedef{#1\leftbracket#1}\wrlabel{#1=#1}}
\def\subseclab#1{\xdef #1{\secsym\the\subsecno}%
\writedef{#1\leftbracket#1}\wrlabel{#1=#1}}
\newwrite\tfile \def\writetoca#1{}
\def\leaderfill{\leaders\hbox to 1em{\hss.\hss}\hfill}
\def\writetoc{\immediate\openout\tfile=toc.tmp
   \def\writetoca##1{{\edef\next{\write\tfile{\noindent ##1
   \string\leaderfill {\noexpand\number\pageno} \par}}\next}}}

%
\catcode`\@=12 
%
\ifx\answ\bigans
\font\titlerm=cmr10 scaled\magstep3 \font\titlerms=cmr7 scaled\magstep3
\font\titlermss=cmr5 scaled\magstep3 \font\titlei=cmmi10 scaled\magstep3
\font\titleis=cmmi7 scaled\magstep3 \font\titleiss=cmmi5 scaled\magstep3
\font\titlesy=cmsy10 scaled\magstep3 \font\titlesys=cmsy7 scaled\magstep3
\font\titlesyss=cmsy5 scaled\magstep3 \font\titleit=cmti10 scaled\magstep3
\else
\font\titlerm=cmr10 scaled\magstep4 \font\titlerms=cmr7 scaled\magstep4
\font\titlermss=cmr5 scaled\magstep4 \font\titlei=cmmi10 scaled\magstep4
\font\titleis=cmmi7 scaled\magstep4 \font\titleiss=cmmi5 scaled\magstep4
\font\titlesy=cmsy10 scaled\magstep4 \font\titlesys=cmsy7 scaled\magstep4
\font\titlesyss=cmsy5 scaled\magstep4 \font\titleit=cmti10 scaled\magstep4
\font\absrm=cmr10 scaled\magstep1 \font\absrms=cmr7 scaled\magstep1
\font\absrmss=cmr5 scaled\magstep1 \font\absi=cmmi10 scaled\magstep1
\font\absis=cmmi7 scaled\magstep1 \font\absiss=cmmi5 scaled\magstep1
\font\abssy=cmsy10 scaled\magstep1 \font\abssys=cmsy7 scaled\magstep1
\font\abssyss=cmsy5 scaled\magstep1 \font\absbf=cmbx10 scaled\magstep1
\skewchar\absi='177 \skewchar\absis='177 \skewchar\absiss='177
\skewchar\abssy='60 \skewchar\abssys='60 \skewchar\abssyss='60
\fi
\skewchar\titlei='177 \skewchar\titleis='177 \skewchar\titleiss='177
\skewchar\titlesy='60 \skewchar\titlesys='60 \skewchar\titlesyss='60
\def\titlefont{\def\rm{\fam0\titlerm}
\textfont0=\titlerm \scriptfont0=\titlerms \scriptscriptfont0=\titlermss
\textfont1=\titlei \scriptfont1=\titleis \scriptscriptfont1=\titleiss
\textfont2=\titlesy \scriptfont2=\titlesys \scriptscriptfont2=\titlesyss
\textfont\itfam=\titleit \def\it{\fam\itfam\titleit} \rm}
%
\def\abstractfont{\def\rm{\fam0\absrm}
\textfont0=\absrm \scriptfont0=\absrms \scriptscriptfont0=\absrmss
\textfont1=\absi \scriptfont1=\absis \scriptscriptfont1=\absiss
\textfont2=\abssy \scriptfont2=\abssys \scriptscriptfont2=\abssyss
\textfont\itfam=\bigit \def\it{\fam\itfam\bigit}
\textfont\bffam=\absbf \def\bf{\fam\bffam\absbf} \rm}
%
\def\tenpoint{\def\rm{\fam0\tenrm}
\textfont0=\tenrm \scriptfont0=\sevenrm \scriptscriptfont0=\fiverm
\textfont1=\teni  \scriptfont1=\seveni  \scriptscriptfont1=\fivei
\textfont2=\tensy \scriptfont2=\sevensy \scriptscriptfont2=\fivesy
\textfont\itfam=\tenit \def\it{\fam\itfam\tenit}
\textfont\bffam=\tenbf \def\bf{\fam\bffam\tenbf} \rm}
%
%

\hyphenation{anom-aly anom-alies coun-ter-term coun-ter-terms}
\def\inv{^{\raise.15ex\hbox{${\scriptscriptstyle -}$}\kern-.40em 1}}

\def\Dsl{\,\raise.15ex\hbox{/}\mkern-13.5mu D} 
\def\dsl{\raise.15ex\hbox{/}\kern-.57em\partial}

\font\bigit=cmti10 scaled \magstep1
\def\lspace{\ifx\answ\bigans{}\else\qquad\fi}
\def\lbspace{\ifx\answ\bigans{}\else\hskip-.2in\fi} 
\def\boxeqn#1{\vcenter{\vbox{\hrule\hbox{\vrule\kern3pt\vbox{\kern3pt
\hbox{${\displaystyle #1}$}\kern3pt}\kern3pt\vrule}\hrule}}}
\def\mbox#1#2{\vcenter{\hrule \hbox{\vrule height#2in
\kern#1in \vrule} \hrule}}    
%

\def\e#1{{\rm e}^{^{\textstyle#1}}}

\def\darr#1{\raise1.5ex\hbox{$\leftrightarrow$}\mkern-16.5mu #1}

\def\roughly#1{\raise.3ex\hbox{$#1$\kern-.75em\lower1ex\hbox{$\sim$}}}

\def\frac#1#2{{#1\over#2}}

\def\journal#1&#2(#3){\unskip, #1~\bf #2 \rm(19#3) }
\def\andjournal#1&#2(#3){\sl #1~\bf #2 \rm (19#3) }

\def\det{{\rm det}}
\def\exp{{\rm exp}}

\catcode`\@=11
\def\slash#1{\mathord{\mathpalette\c@ncel{#1}}}
\overfullrule=0pt
\def\steepslash{\c@ncel}
\def\frac#1#2{{#1\over #2}}

\def\e{\epsilon}

\def\p {\partial}
\def\Z{{\bf Z}}
\def\C{{\bf C}}

\def\R{{\bf R}}
\def\inbar{\,\vrule height1.5ex width.4pt depth0pt}
\def\IB{\relax{\rm I\kern-.18em B}}
\def\IC{\relax\hbox{$\inbar\kern-.3em{\rm C}$}}
\def\IP{\relax{\rm I\kern-.18em P}}
\def\IR{\relax{\rm I\kern-.18em R}}
\def\IZ{\relax\ifmmode\mathchoice
{\hbox{Z\kern-.4em Z}}{\hbox{Z\kern-.4em Z}}
{\lower.9pt\hbox{Z\kern-.4em Z}}
{\lower1.2pt\hbox{Z\kern-.4em Z}}\else{Z\kern-.4em Z}\fi}

\catcode`\@=12

\def\npb#1(#2)#3{{ Nucl. Phys. }{\bf B#1} (#2) #3}
\def\plb#1(#2)#3{{ Phys. Lett. }{\bf #1B} (#2) #3}
\def\pla#1(#2)#3{{ Phys. Lett. }{\bf #1A} (#2) #3}
\def\mpla#1(#2)#3{{ Mod. Phys. Lett. }{\bf A#1} (#2) #3}
\def\ijmpa#1(#2)#3{{ Int. J. Mod. Phys. }{\bf A#1} (#2) #3}
\def\cmp#1(#2)#3{{ Comm. Math. Phys. }{\bf #1} (#2) #3}
\def\cqg#1(#2)#3{{ Class. Quantum Grav. }{\bf #1} (#2) #3}
\def\jmp#1(#2)#3{{ J. Math. Phys. }{\bf #1} (#2) #3}
\def\anp#1(#2)#3{{ Ann. Phys. }{\bf #1} (#2) #3}

\def\a{\alpha}
\def\b{\beta}
\def\d{\delta}
\def\g{\gamma}
\def\o{\omega}

\def\e{\epsilon}

\def\p1{\phi_1}
\def\p2{\phi_2}

\def\({\lbrack}
\def\){\rbrack}
\def\mat{\matrix}
\def\pmat{\pmatrix}

\Title{KA-THEP-03/92}{TUM-TH-143-92}
{Considerations of One-Modulus Calabi-Yau
Compactifications:}
{Picard-Fuchs Equations, K\"ahler Potentials and Mirror Maps}
\centerline{Albrecht Klemm\footnote{$^*$}{email:aklemm@physik.tu-muenchen.de}
\footnote{$\dag$}{Supported by the Deutsche Forschungsgemeinschaft.}}
\medskip\centerline{Physik Department T30, Technische Universit\"at
M\"unchen}
\centerline{James-Franck-Stra\ss e, D - 8046 Garching, FRG}
\bigskip
\centerline{Stefan Theisen\footnote{$^{**}$}{email: BE01@DKAUNI2}}
\medskip\centerline{Institut f\"ur Theoretische Physik,
Universit\"at Karlsruhe}
\centerline{Kaiserstr. 12, D - 7500 Karslruhe, FRG}
\vskip .5in

{\noindent {\bf Abstract:} We consider Calabi-Yau compactifications with one
K\"ahler modulus. Following the method of Candelas et al.
we use the mirror hypothesis to solve the quantum theory exactly
in dependence of this modulus by performing the calculation for
the corresponding complex structure deformation on the mirror manifold.
Here the information is accessible by techniques of classical geometry.
It is encoded in the Picard-Fuchs differential equation
which has to be supplemented by requirements on the global properties
of its solutions.}

\Date{april/92}

\footline{\hss\tenrm--\folio\--\hss}

\newsec{Introduction}
One of the outstanding problems in string theory is to close the
gap between the formal description and classification
of string vacua and their possible role in a realistic
description of particle physics. Even if one finds a model
with the desired particle content and gauge symmetry, one is
still confronted with the problem of computing the couplings,
which determine masses, mixing angles, patterns of symmetry
breaking etc. These couplings will depend on the moduli of
the string model, which, in the conformal field theory language
correspond to the exactly marginal operators, or, in the
Calabi-Yau context, to the (1,1) and (2,1) forms, which
describe the deformation of the K\"ahler class and the
complex structure, respectively. In (2,2) compactifications,
which are the ones which have been most intensively studied
to date, the two types of moduli are related by world sheet
supersymmetry to the matter fields, which transform as
$\overline{27}$ and $27$ of $E_6$. In a low energy effective field
theory description, which includes all the light states, but
having integrated out all heavy ($>m_{\rm Planck}$) string modes,
the moduli appear as massless neutral scalar fields with
 vanishing potential. This entails that the
strength of the couplings, such
as the Yukawa couplings, which do depend on the moduli, are
undetermined. Only if the vacuum expectation value of the
moduli fields is fixed by a non-perturbative potential do the
couplings take fixed values, which could then be compared
with experiment.

Generic string models are believed to possess duality
symmetry\ref\dualgen{K. Kikkawa and M. Yamasaki, \plb149(1984)357;
N. Sakai and I. Senda, Prog. Theor. Phys. {\bf 75} (1984) 692;
V. P. Nair, A. Shapere, A. Strominger and F. Wilczek,
\npb287(1987)402; A. Giveon, E. Rabinovici and G. Veneziano,
\npb322(1989)167.},
which is a discrete symmetry on moduli space
that leaves the spectrum as well as the
interactions invariant and whose origin is tied to the fact that strings
are one-dimensional extended objects. This symmetry has been explicitly
found in simple models, such as the compactification on tori and
their orbifolds
\ref\orbdual{  M. Dine, P. Huet and N. Seiberg, \npb322(1989)301;
J.Lauer, J. Mas and H. P. Nilles, \plb226(1989)251;
M. Spali\'nski, \plb275(1992)47.},
but more recently for a specific Calabi-Yau
compactification
\ref\cdgp{P. Candelas, X. de la Ossa, P. Green and
L. Parkes, \npb359(1991)21.}.

On the effective field theory level this string specific
symmetry is manifest insofar as the Lagrangian must be
invariant\ref\fslt{S. Ferrara, D. L\"ust, A. Shapere and S. Theisen,
\plb225(1989)363.}.
This has the important consequence that the moduli
dependent couplings must have definite transformation properties
under transformations of the duality
group\footnote{$^*$}{For the simplest case where the duality
group is just the modular group $SL(2,\Z)$, they are
modular forms.}. A possible non-perturbative potential for
the moduli fields must also respect this symmetry.

The problem then consists of first computing the moduli dependence of
the Yukawa couplings, to find candidates for the potential
for the moduli and then to look for its minimum, thus fixing the
value of the Yukawa couplings.

This is a formidable task to perform for a generic string model, and has
thus far only partially been done for the few simple models mentioned above.
For the toroidal orbifold examples this is not too difficult
if one restricts oneself to the {\it untwisted moduli}. The dependence of the
couplings on the untwisted moduli can be calculated in conformal field
theory\ref\orbicoup{
L. Dixon, D. Friedan, E. Martinec and S. Shenker,
\npb282(1987)13;
S. Hamidi and C. Vafa, \npb279(1987)465;
J. Lauer, J. Mas and H. P. Nilles, \npb351(1991)362.}
and the duality group of the orbifold is the subgroup
of the torus duality group\dualgen\
which is compatible
with the discrete group by which the orbifold is defined\orbdual .
In simple cases also the corresponding modular functions are known.
The situation for Calabi-Yau manifolds is more involved, partially because one
knows the conformal field theory explicitly only at special points
in moduli space, where the models coincide with the ones of
Gepner
\ref\gep{D. Gepner,\npb296(1988)757;
{\it N=2 String Theory};
Lectures at the Trieste spring school
on superstrings (1989), PUPT-1121.}.
Given the recent results of \ref\iblu{L. Ibanez and
D. L\"ust, {\it Duality Anomaly Cancellation, Minimal String Unification
and the Effective Low-Energy Lagrangian of 4-D Strings}, preprint
CERN-TH-6380-92.}
which exclude most simple toroidal orbifold
models as viable candidate string vacua, one is harder pressed
to develop tools to do explicit computations for more complicated
compactifications. An important step in this direction has
been done in \cdgp.

Above we have already mentioned the two different kinds of moduli. The
moduli space is a direct product with a separate factor for the
(1,1) and the (2,1) moduli\ref\dkletal{N. Seiberg, \npb303(1988)286;
S. Cecotti, S. Ferrara and L. Girardello, Int. J. Mod. Phys.
{\bf A4} (1989) 2475; L. Dixon, V. Kaplunovsky and J. Louis,
\npb329(1990)27.}.
Since in each case, due to the fact that
the same (2,2) superconformal field theory could have been used
to compactify the type II rather than the heterotic string with
the former leading to $N=2$ space-time supersymmetry, the moduli
space is of special K\"ahler type\ref\special{For reviews see
S. Ferrara, Mod. Phys. Lett. {\bf A6} (1991) 2175;
S. Ferrara and S. Theisen, in Proceeedings of the Hellenic Summer
School 1989, World Scientific; A. Strominger \cmp133(1990)163;
P. Candelas and X. de la Ossa, \npb355(1991)455.}.
This means in particular
that the K\"ahler metric of moduli space is completely
determined in terms of two holomorphic functions, the prepotentials,
one for each type of moduli. The Yukawa couplings are give
by the third derivatives of the prepotentials with respect
to the moduli. This entails that they do not mix
the two sets of moduli and their corresponding matter fields;
i.e. the Yukawa couplings of the $\overline{27}'s$ of $E_6$ only
depend on the K\"ahler moduli and the couplings of the
$27's$ depend only on the complex structure moduli.
Whereas the former acquire contributions from instantons, the latter
do not\ref\digr{J. Distler and B. Greene, \npb309(1988)295.}
and are thus in general easier to compute.

On the conformal field theory level the $27's$
and $\overline{27}'s$ of $E_6$, and by world-sheet supersymmetry
the two types of moduli, can be simply interchanged by flipping
the relative sign of the left and right $U(1)$ charges of
the (2,2) superconformal algebra
\ref\grpl{B. Greene and Plesser,\npb338(1990)15.}.
On the geometrical level this corresponds to an interchange
of the Hodge numbers
$h_{1,1}$ and $h_{2,1}$ and thus of a change of sign of the
Euler number. This so called mirror map relates topologically
distinct Calabi-Yau spaces. The mirror hypothesis states that
the prepotentials for the different types of moduli are
interchanged on the manifold and its mirror.
Mirror symmetry thus allows
one to get the instanton corrected couplings for the
(1,1) forms on a given Calabi-Yau manifold $M$ from the
couplings of the (2,1) forms on its mirror $M^\prime$, which have no
instanton corrections. Following Candelas et al. we will
use this strategy to compute the prepotential,
K\"ahler potential and Yukawa couplings for the four
Calabi-Yau spaces with $h_{1,1}=1$
in the lists of
refs.\ref\cls{P. Candelas, M. Lynker and R. Schimmrigk,
\npb341(1990)383.},
\ref\fkss{J. Fuchs, A. Klemm, Ch. Scheich and
M.G. Schmidt,\plb232(1989)317,\anp204(1990)317.}.
Similar methods have been used in ref.\ref\morrison{D.R. Morrison,
{\it Picard-Fuchs equations and mirror maps for hypersurfaces},
preprint DUK-M-91-14.} to compute the instanton numbers on
these spaces. In one of the four cases our results differ.


The paper is organized as follows: in the next section we introduce the
models which we will discuss. We then (sect. 3) set up the
period equations and discuss their solutions, including
their monodromy properties. In section four we construct
a basis for the solutions on which the monodromy transformations
are integer symplectic and in terms of which the prepotential
can be easily written down. We compute the K\"ahler metric and the
invariant Yukawa couplings. In section five we perform the mirror
map. In the conclusions we make some comments on the
modular group and a candidate for the non-perturbative
potential for the modulus. To a large extent our exposition
will follow ref.\cdgp\ .

\newsec{The models}
The simplest Calabi-Yau models are described as the vanishing
locus of a quasi-homogeneous polynomial in five variables
of the Fermat type
$W_0=\sum_{i=0}^4 x_i^{n_i}=0$
which gives the embedding of the three (complex)dimensional
manifold in weighted projective 4 - space $\IP^4$.
Vanishing of the first Chern class and absence of singularities, whose
resolution would introduce new $(1,1)$-forms,
imposes severe restrictions on the $n_i$, leaving only
four manifolds which all have $h_{1,1}=1$, i.e. possess
only one K\"ahler modulus. Characterizing these models
by the integer $k$, which is defined to be the smallest
common multiple of the $n_i$, they are (the relative
factors are chosen for later convenience)
\eqn\models{\eqalign{
  k=5~~:\qquad\quad M=\{x_i\in \IP(1,1,1,1,1)\,\, |\,\,
W_0&=x_0^5+x_1^5+x_2^5+x_3^5+x_4^5=0\}\cr
  k=6~~:\qquad\quad M=\{x_i\in \IP(2,1,1,1,1)\,\,|\,\,
W_0&=2x_0^3+x_1^6+x_2^6+x_3^6+x_4^6=0\}\cr
  k=8~~:\qquad\quad M=\{x_i\in \IP(4,1,1,1,1)\,\, |\,\,
W_0&=4x_0^2+x_1^8+x_2^8+x_3^8+x_4^8=0\}\cr
  k=10:\qquad\quad M=\{x_i\in \IP(5,2,1,1,1)\,\, |\,\,
W_0&=5x_0^2+2x_1^5+x_2^{10}+x_3^{10}+x_4^{10}=0\}\cr}}
These manifolds have first been found by Strominger and
Witten \ref\stwi{A. Strominger and E. Witten,\cmp101(1985)341.}.
They belong to a class of superstring compactifications whose
internal space can be described at a special point of its moduli space
by tensor products of minimal $(2,2)$ super conformal theories with
$A$-type modular invariants \ref\gep{D. Gepner,\npb296(1988)757;
{\it N=2 String Theory};
Lectures at the Trieste spring school
on superstrings (1989), PUPT-1121.}, often called Gepner models.
A survey of this class with $A$-$D$-$E$-type modular invariants can be
found in\footnote{$^*$}{Recently more
complicated models with $h_{1,1}=1$
have been constructed \ref\klsch{A. Klemm and R. Schimmrigk,
{\it  Landau--Ginzburg String Vacua}, preprint CERN-TH-6459/92,
TUM-TP-142/92.}. We will however not consider them here.}
\ref\lr{A. L\"utken and G. Ross, \plb213(1998)152.},\fkss\ .
The complex structure moduli of the models \models\  can be represented
by  those elements in the polynomial ring
${\cal R}={\C\(x_i\)\over dW_0}$ which are of the same
degree as $W_0$. They correspond to the marginal deformations
of the associated $(2,2)$ conformal field theory.
One finds $h_{2,1}=101, 103, 149, 145$
for $k=5,6,8,10$ respectively, corresponding to the
Euler numbers $\chi=2(h_{1,1}-h_{2,1})=-200, -204, -296, -288$.

One can now consider orbifolds of these spaces by dividing out
discrete isometries which will in general not act freely.
If the isometry acts like a subgroup of $SU(3)$ the possible singularities
can be resolved such that the resulting space in again of Calabi-Yau type
as it was conjectured in \ref\dhvw{L. Dixon, J. A. Harvey, C. Vafa and
E. Witten,\npb261(1985)678.}. This process changes the Euler number and
thus also the Hodge numbers $h_{2,1}$ and $h_{1,1}$.
For our investigation the most important groups
are the ones which lead to the mirror configuration.
They are generated by multiplying the $\IP^4$ coordinates with phases:
$x_i\rightarrow x_i\,\exp[{2 \pi i\over n_i} \cdot r_i]$. We abbreviate
generators as $g=(r_0,\ldots,r_5)$. The condition for $g\in SU(3)$
reads simply $\sum[{r_i\over n_i}]=0\,\, mod\,\, 1$.
In Appendix A we list all possible Hodge numbers which can be
obtained from our models by dividing out all subgroups of the full phase
symmetry group $G$\footnote{$^{**}$}{Here we do not consider the
moding of (nonabelian) permutation symmetries
\ref\asluro{P.S. Aspinwall, A. L\"utken
and G. Ross,\plb241(1990)373.},\ref\klsch{A. Klemm and M.G. Schmidt,
\plb245(1990)53;
J. Fuchs, A. Klemm and M. G. Schmidt,\anp214(1992)221.}.} with $G\in SU(3)$.
The topological data can be obtained
directly by a tedious calculations on
the manifold, using the properties of the resolutions
\ref\fujiki{A. Fuijki, Publ. RIMS, Kyoto Univ. {\bf 10} (1974) 293.},
\ref\roya{S. Roan and S.T. Yau, Acta Math. Sinica {\bf 3} (1987) 256.},
\ref\klemm{A. Klemm, {\it Stringkompaktifizierung durch
Tensorprodukte minimaler n=2 superkonformer Feldtheorien
und ihre geometrische Interpretation}, Ph.D. Dissertation,
Heidelberg University 1990.} but more easily by using
the methods of twisting the Landau-Ginzburg models
\ref\vafa{C. Vafa,\mpla12(1989)1169,\mpla17(1989)1615.}
or by examining the
massless spectrum of the corresponding Gepner model
\ref\zog{P. Zoglin,\plb218(1989)444.}, \klsch .
The latter method gives the full
partition function of the twisted theory.
For the Fermat  cases  ($A$-type modular
invariants) the Hodge numbers obtained this way were veryfied in
\ref\roan{S.S. Roan, Int. Jour. Math. {\bf 1} (1990) 211; see also
{\it Topological Couplings of Calabi-Yau-Orbifolds},
Preprint MPI/92/22.}
by explicit construction of the geometrical resolution.
In the orginal models we consider, all $(2,1)$-forms can be described by
deformations of the defining polynomial. The orbifoldisation projects onto
those $(2,1)$ forms which are invariant,
their number is given in parantheses in our tables,
but also introduces new ones in the twisted sectors,
which cannot be represented as deformations of
$W_0$. For a geometric representation of them see
\ref\bgh{P. Berglund, B. R. Greene, T. H\"ubsch, {\it Classical vs.
Landau-Ginzburg Geometry of Compactifications.}, preprint CERN-TH-3681-92}.
In all four cases, dividing out the
whole of $G$, one finds that only one possible deformation
survives, thus giving $h_{2,1}=1$.
The resulting quotient models always appear in mirror
pairs with $h_{2,1}$ and $h_{1,1}$ interchanged.
See also \grpl,
\asluro, \klemm\
for the cases
$k=5$\footnote{$^{\dag}$}{For the $k=10$ case the list given in
\grpl \ is incomplete.};
the case $k=8$ was discussed in \klemm .

The actual order of the group action, which is very important in order to
normalize our period integrals, is safely investigated on the
Calabi-Yau manifold. In the $k=8$ case e.g. the generators
$g_1=(0\,1\,0\,0\,7)$,  $g_2=(0\,1\,0\,7\,0)$, $g_3=(0\, 1\,7\,0\,0)$
naively seem to generate a $Z_8\times Z_8\times Z_8$
group. However, the element $g'= (2\, g_1)\times(2\, g_2)
\times (2\, g_3)\simeq
(0,-2,-2,-2,-2)$ generates a $Z_4$ subgroup which operates trivial
on the coordinates, because of the equivalence relation
of the $\IP(4,1,1,1,1)$ we have $(z_1,-i\,z_2,-i\, z_3,-i\, z_4,-i\,
z_5)\simeq (z_1, z_2, z_3,z_4, z_5)$.
Hence the actual group acts as a
$Z_8\times Z_8\times Z_2$. In the Gepner- or Landau-Ginzburg model
language this corresponds to the fact that the afore
mentioned $Z_4$ acts a subgroup of the group by which is
divided out in order to implement the GSO projection. The
actual groups which generate the mirror configurations are thus
$G=Z_5^3,\, Z_3\times Z_6^2,\,
Z_8^2\times Z_2$ and $Z_{10}^2$ with ${\rm Ord}(G)=5^3,\,3\cdot 6^2,\,
2\cdot 8^2$ and $10^2$
in the four cases, respectively. We observe that
$({\rm Ord}G)\prod_{i=0}^4\nu_i=k^3$,
where the $\nu_i={k\over n_i}$ are the weights of $\IP^4$
coordinates which satisfy $\sum_{i=0}^4\nu_i=k$.
We then get for the deformed polynomials
\eqn\defpol{W=W_0-k\alpha x_0 x_1 x_2 x_3 x_4}
where the perturbation can always be cast into the indicated
form by the use of the equations of motion $\partial_i W=0$.
The constants in \models\ and \defpol\ have been chosen
such that $\alpha^k=1$ are nodes in all four cases.
The elements of the rings ${\cal R}$ are now also restricted
by the discrete symmetries. In the case where one divides
out all of $G$, ${\cal R}$ consists of only the elements
$(x_1 x_2 x_3 x_4 x_5)^\lambda$ with $\lambda=0,1,2,3$.
Besides having nodes at $\a^k=1$ the manifolds become singular
at $\a\to\infty$.

\newsec{The Picard-Fuchs equations and their solutions}
To summarize the results of refs.\ref\lesmwa{W. Lerche, D.-J. Smit
and N. Warner, \npb372(1992)87.},
\ref\feca{Cadavid and S. Ferrara,\plb267(1991)193;
S. Ferrara and J. Louis, \plb278(1992)240;
A. Ceresole, R. D'Auria, S. Ferrara, W. Lerche
and J. Louis, {\it Picard-Fuchs equations and special geometry};
preprint CERN-TH.6441/92.},
\ref\klsmth{A. Klemm, M.G. Schmidt and S. Theisen,
{\it Correlation functions for topological Landau-Ginzburg
models with $c\leq3$}, preprint TUM-TP-129/91; Int. J. Mod. Phys.
{\bf A}, to be published.}
the Picard-Fuchs or period equations
are differential equations satisfied
by the expression $w=\int{q(\a)\over W(\a)}$ where the integral
is in the embedding space and
allows integration by parts with respect to the coordinates
of  $\IP^4$.
The holomorphic
function $q(\a)$ reflects the gauge freedom
in the definition of the holomorphic three-form. For the purpose
of deriving the period equation, it is most convenient to
set $q(\a)=1$. Differentiating $\lambda$ times with respect to $\a$
produces terms of the form $\int{(x_1 x_2 x_3 x_4 x_5)^\lambda
\over W^{\lambda+1}(\a)}$. The $\lambda=4$ term, which is the first
to produce an integrand whose numerator is no longer in the
Ring ${\cal R}$, can be expressed, using the equations of motion
$\partial W/\partial x_i=0$
and integration by parts, in terms of lower derivatives.
The computation is straightforward and
produces\footnote{$^*$}{Manifolds having the mirror Hodge
diamond w.r.t. the models in \models\ can be found in the lists of
\cls , \klsch . As has been checked in a few examples\ref\tk{A. Klemm
and S. Theisen; unpublished.} the Picard-Fuchs
equations for the only complex structure deformation here
turn out to be the same as in (3.1).}
\eqn\pereq{\eqalign{
k=5:\,\,\quad &(1-\a^5)\,w^{(iv)}-10\,\a^4\, w^{\prime\prime\prime}\,
-25\,\a^3\, w^{\prime\prime}-15\,\a^2\, w^{\prime}\,- \a\,w=0\cr
k=6:\,\,\quad &\a^2(1-\a^{6})\,w^{(iv)}-2\a(1+5\a^6)\,w^{\prime\prime\prime}\,
+(2-25\a^6)\,w^{\prime\prime}-15\a^5\,w^\prime-\a^4\,w=0\cr
k=8:\,\,\quad &\a^3(1-\a^8)\,w^{(iv)}-\a^2(6+10\a^8)\,w^{\prime\prime\prime}
+5\a(3-5\a^8)\,w^{\prime\prime}-15(1+\a^8)\,w^\prime-\a^7\,w=0\cr
k=10:\quad &\a^3(1-\a^{10})\,w^{(iv)}
-10\a^2(1+\a^{10})\,w^{\prime\prime\prime}\cr
&\qquad\qquad
+5\a(7-5\a^{10})\,w^{\prime\prime}-5(7+3\a^{10})\,w^\prime-\a^9\,w=0\cr}}
A fundamental system of solutions may be obtained
following the method of Froebenius for ordinary differential
equations with regular singular points \ref\ince{E.L. Ince,
{\it Ordinary differential equations}, Dover 1956.}
which are here
$\a=0$, $\a=\infty$ and $\a^k=1$.
The solutions of the indicial equations
at the three singular points are
$\rho=(0,1,2,3)_{k=5},~(0,1,3,4)_{k=6},~(0,2,4,6)_{k=8},~
(0,2,6,8)_{k=10}$ for $\a=0$, $\rho=(0,1_2,2)$ for $\a^k=1$
and $\rho=0_4$ for $\a=\infty$. The subscripts denote the
multiplicities of the solutions. It follows from the
general theory that at $\a=\infty$ there is one solution
given as a pure power series and three containing logarithms
(with powers 1,2 and 3, respectively). At $\a=0$, all four
solutions are pure power series as one sees e.g. by
noting that we can rewrite the differential equation
in terms of the variables $\a^k$, for which the solutions of
the indicial equation would no longer differ by integers.
The point $\a=1$ needs some care\footnote{$^{**}$}{The other solutions
of $\a^k=1$ are treated similarly.}. There is one power series
solution with index $\rho=2$ and at least one logarithmic
solution for $\rho=1$.
Making a power series ansatz for $\rho=0$ one finds that the
first three coefficients are arbitrary which means that there is
one power series solution for each $\rho$. One also easily
checks that in the second solution to $\rho=1$ the
logarithm is multiplied by a linear combination of the power
series solutions with indices 1 and 2.
To summarize, the periods of the manifolds have logarithmic
singularities at the values of $\a$ corresponding to the
node ($\a^k=1$) and to the singular manifold ($\a=\infty$).
We will thus get non-trivial monodromy about these points.

The power series solution around $\a=\infty$ is easy to find
by making a general ansatz and solving the recursion relation
for the coefficients. One finds
\eqn\solinfty{w_0(\a)={1\over\a}\,\sum_{m=0}^\infty{(km)!\over
\prod_{i=0}^4\left(\nu_i\,m\right)!}(\g\a)^{-km}}
where $\gamma=k\prod_{i=0}^4\left(\nu_i\right)^{-\nu_i/k}$.
The other solutions around $\a=\infty$ contain logarithms.
We will find them below. For later convenience we will
redefine the periods to get rid of the factor
$\a^{-1}$. This corresponds to the gauge transformation
$q(\a)=1\to q(\a)\propto\a$. This only affects the exponents
at $\a=0$ which are shifted by +1. To get the solutions around
$\a=0$ we first analytically continue above solution.
If $0\leq{\rm arg}\,\a<{2\pi\over k}$, we have
\eqn\barnes{w_0(\a)={1\over 2\pi i}\int_C ds\,
{\Gamma(1+ks)\over\prod_{i=0}^4\Gamma(1+\nu_i s)}\,
{\pi e^{i\pi s}\over \sin(\pi s)}\,(\g\a)^{-ks}\,.}
For $|\a|>1$ we get a convergent expression if we choose the contour
to enclose the points $\a=n=0,1,2\dots$ which are zeros
of $\sin(\pi s)$, in a clockwise direction
and we recover
\solinfty.
For $|\a|<1$ we close the contour as to enclose
the points $-n/k$, $n=1,2\dots$ which are the poles of
$\Gamma(ks+1)$ with residues $(-1)^{n+1}/\left(k\Gamma(n)\right)$.
One finds
\eqn\solzero{w_0(\a)=-{\pi\over k}\sum_{n=1}^{\infty}
{1\over\Gamma(n)
\prod_{i=0}^4\Gamma\left(1-{n\over k}\nu_i\right)}\,
{e^{i{\pi\over k}(k-1)n}\over \sin\left({\pi n\over k}\right)}\,
(\gamma\a)^n\,.}
All solutions at $\a=0$ are given by power series and one
readily sees that the functions $w_j(\a):=w_0(\b^j\a),~
(j=0,1,\dots,k-1)$ with $\b=e^{{2\pi i\over k}}$ are also solutions.
These are however not all linearly independent. There are
$k-4$ linear relations.
One possible choice is $\sum_{j=0}^4 w_j=0$ for $k=5$,
$w_0+w_2+w_4=
w_1+w_3+w_5=0$ for $k=6$,
$w_{i}+w_{i+4}=0,~(i=0,1,2,3)$ for $k=8$,
$w_i+w_{i+5}=0,~(i=0,1,2,3,4)$
and $w_0+w_2+w_3+w_4+w_5=0$
for $k=10$.
Below we will use the functions $w_0,\,w_1,\,w_2$ and $w_{k-1}$
as a basis.

To get all the solutions for $\a=\infty$ we first express the
$w_j$ in terms of four linearly independent solutions which
we then analytically continue from $|\a|<1$ to $|\a|>1$.
Changing the summation index to $n=kN+l$, $N=0,1,2\dots,
{}~l=1,\dots,k-1$, one obtains
\eqn\solszero{w_j(\a)=-{1\over 16 k\pi^4}
\sum_l \b^{jl}\, {\prod_{i=0}^4(\b^{l\nu_i}-1)\over{\b^l -1}}\,
\tilde w_l(\a)}
with
\eqn\wtilde{\eqalign{\tilde w_l(\a)&=\sum_{N=0}^\infty
{\prod_{i=0}^4
\Gamma\left(\nu_i\left(N+{l\over k}\right)\right)\over
\Gamma(kN+l)}\,(\gamma\a)^{kN+l}\cr
&=-\int_C{ds\over e^{2\pi i s}-1}\,
{\prod_{i=0}^4\Gamma\left(\nu_i\left(s+{l\over k}\right)\right)\over
\Gamma(ks+l)}\,(\gamma\a)^{ks+l}\cr}},
where to recover \solszero\ one has to choose the contour
as to enclose the poles of $(e^{2\pi is}-1)^{-1}$. Note that in
eq.\solszero\ there are only four non-zero terms in each sum,
namely for $l=(1,2,3,4)_{k=5},~l=(1,2,4,5)_{k=6},~
l=(1,3,5,7)_{k=8}$ and $l=(1,3,7,9)_{k=10}$.
To get the solutions for $|\a|>1$ one chooses the contour
to surround the poles of the second factor of the integrand
which has quadruple poles at $s=-N-{l\over k}$ for $N=0,1,\dots$.
Evaluating the residues at the poles is straightforward.
Expanding $\a^{ks+l}$ around the poles produces up to three
powers of $\log\a$. Collecting terms one finds
\eqn\solsinfty{w_j(\a)=-{1\over(2\pi i)^3}
{1\over\prod_{i=0}^4\nu_i}\sum_{r=0}^3\log^r(\g\a)\sum_{N=0}^\infty
{(kN)!\over\prod_{i=0}^4(\nu_iN)!}\, b_{jrN}\, (\g\a)^{-kN}}
where
$$
\eqalign{
b_{j0N}=&(2\pi i)^3\left(S_{j4}-\prod_{i=0}^4\nu_i\right)
+(2\pi i)^2\left(2\pi i+k\phi(N)\right)S_{j3}\cr
&+{2\pi i\over 24}\left((2\pi i)^2\Bigl(4+k^2-\sum_{i=0}^4\nu_i^2\Bigr)
+12(2\pi i)k\phi(N)+12\left(k^2\phi^2(N)-k\phi^\prime(N)\right)
\right)S_{j2}\cr
&+{k\over 24}\left((2\pi i)^2\Bigl(k^2-\sum_{i=0}^4\nu_i^2\Bigr)
\phi(N)+4k^2\phi^3(N)-12k\phi(N)\phi^\prime(N)
+4\phi^{\prime\prime}(N)\right)S_{j1}\cr
\noalign{\smallskip}
b_{j1N}=&k(2\pi i)^2 S_{j3}+{k\over 2}(2\pi i)
\Bigl((2\pi i)+2k\phi(N)\Bigr)S_{j2}\cr
&\qquad+{k\over 24}\left((2\pi i)^2\Bigl(k^2-\sum_{i=0}^4
\nu_i^2\Bigr)+12k^2\phi^2(N)-12k\phi^\prime(N)\right)S_{j1}\cr
\noalign{\smallskip}
b_{j2N}=&{k^2\over2}\Bigl((2\pi i)S_{j2}+k\phi(N)S_{j1}\Bigr)\cr
\noalign{\smallskip}
b_{j3N}=&{k^3\over 6}S_{j1}\cr}
$$
Here we have defined
$$
\eqalign{\phi_k(N)&={1\over k}\sum_{i=0}^4 \nu_i
\psi(1+\nu_i N)-\psi(1+kN)\,,
\qquad\psi(x)=d\log \Gamma(x)/dx\cr
S_{jm}&=\sum_{l=0}^{k-1}\b^{l(j+1)}{\prod_{i=0}^4(\b^{l\nu_i }-1)\over
(\b^l-1)^{m+1}}\cr}
$$
We now discuss the monodromy of the solutions. Under the
transformation $\a\to\b\a$, the solutions behave as
$w_j(\b\a)\to w_{j+1}(\a)$ and the corresponding
monodromy matrix $a$ is cyclic of order $k$.
{}From our discussion above of the solutions of the period
equations around $\a=1$ we conclude that the $w_j$, when
continued to the region $|\a-1|<1$, must be of the form
\eqn\soleins{w_j(\a)={1\over 2\pi i}c_j
\tilde w(\a)\log(\a-1)\,+\,{\rm regular}}
where $\tilde w(\a)$ is a particular combination
of the power series solutions with indices 1 and 2.
In fact, we can express it in terms of the $w_j$ as follows\cdgp:
\eqn\wtilde{\tilde w(\a)=-{1\over c_1}\left(w_1(\a)-w_0(\a)\right).}
Eq.\soleins\ then says that under
$(\a-1)\to e^{2\pi i}(\a-1)$ the solutions transform as
\eqn\transsoleins{w_j(\a)\to w_j(\a)
-{c_j\over c_1}\left(w_1(\a)-w_0(\a)\right)\,.}
We thus find that the monodromy matrix $t$ around $\a=1$
is not cyclic of any finite order.
To determine the coefficients $c_j$ we normalize
$\tilde w(\a)=\mu(\a-1)+O\left((\a-1)^2\right)$
such that $\mu c_1=1$.
$\mu c_j$ is then the coefficient of
the logarithm of $d w_j/d\a$ as $\a\to 1$. Using the
explicit expressions for $w_j$ (eq.\solszero\ ) we find
$\mu=-{1\over 2\pi i}k^{3/2}\left(\prod_{i=0}^4\nu_i\right)^{-1/2}$
and
\eqn\cj{c_j={1\over k}S_{j-1,0}\,.}
The $c_j$ are all integers.
In all cases $c_0=c_1$, which is necessary for \wtilde\
to be free from logarithms.
Monodromy transformations about the points $\b^l$ follow simply
{}From a composition of the transformations $a$ and $t$.
The monodromy about $\a=\b^l$ is then $a^{-l}ta^l$.\footnote{$^*$}
{Note that if one represents, as we do, the monodromy group on the
fundamental system by matrix multiplication from the left,
the matrices form an anti-homomorphism of the monodromy group.}
Finally, the monodromy matrix $s$ around
$\a=\infty$ follows from the fact that the product
of the monodromy matrices around all singular points must be
the identity and that the monodromy around $\a=0$ is trivial.
Thus $s=\left((at)^{-1}\right)^k$.

\newsec{The periods in a symplectic basis and their monodromy}
In order to get the prepotential
${\cal G}$ from the solutions of the
period equations, we look for a basis in which the monodromy
acts as $Sp(4;\Z)$ transformations.
${\cal G}$ is then given as ${\cal G}={1\over 2}{\cal G}_a z^a$,
where the periods ${\cal G}_a$ and $z^a$ ($a=1,\dots,h_{2,1}$)
are the integrals of the holomorphic three form over a
symplectic basis for the homology group $H_3(\Z)$.
${\cal G}$ is homogeneous of degree two in the homogeneous
coordinates $z^a$ and ${\cal G}_a=\partial{\cal G}/\partial z^a$.
In Appendix B we directly identify two of the solutions
of the Picard-Fuchs equation with ${\cal G}_2$ and $z^2$. With
the gauge choice $q(\a)=k\a$ we find
${\cal G}_2(\a)=\lambda_1 w_0(\a)$ with
$\lambda_1={(2\pi i)^3\over{\rm Ord} G}$.
$z^2(\a)$ is a solution which, around $\a=1$, is a pure
power series with index 1, proportional to $\tilde w(\a)$
in eq.\transsoleins\ and is given, to leading order as
$z^2(\a)=\lambda_2(\a-1)+O\left((\a-1)^2\right)$ with
$\lambda_2={4\pi^2\over k^{3/2}}\left(\prod\nu_i\right)^{1/2}$.
Then $\lambda_2/\lambda_1=\mu$ and
the monodromy coefficients
$c_j$ become integers.

We now define two period vectors\cdgp,
$\vec\Pi^\prime=({\cal G}_1,{\cal G}_2,z^1,z^2)^{\rm T}$ and\break
$\vec w=-{(2\pi i)^3\over{\rm Ord}G} (w_2,w_1,w_0,w_{k-1})^{\rm T}$.
Since they both represent a fundamental set of solutions of the
period equations, they must be related by a linear transformation
which we call $m$.
{}From the above identification of two of the components
of $\vec\Pi$ with elements of $\vec w$, we already know the
second and last row of the matrix $m$.
Since the cycles dual to the periods ${\cal G}_1$ and $z^1$
are remote from the node at $\a=1$, they must be free from
logarithms. This gives a constraint on the first and third
row of $m$. We then fix the remaining six components
of $m$ by requiring that under $\a\to\b\a$, which was
represented on the basis $w_i$ by the matrix $a$, the vector
$\vec\Pi^\prime$ transforms under an integer
symplectic transformation $A$;
i.e. we determine $m$ such that $A=m\,a\,m^{-1}\in SP(4;\Z)$.
\footnote{$^*$}{A matrix $M$ is symplectic if it satisfies
$M\Sigma M^{\rm T}=\Sigma$, where $\Sigma=\pmatrix{\phantom{-}
{\bf 0}&{\bf 1}\cr {\bf -1}&{\bf 0}\cr}$ is the symplectic
metric.}
This does not yet fix the matrix $m$ uniquely. Having defined
the periods ${\cal G}_2$ and $z^2$, the remaining two
can only be fixed by the above requirement up to a
$SP(2;\Z)\subset SP(4;\Z)$ transformation. With a suitable
choice we obtain the results given below. We have also
recorded the transformation matrices $S=m\,s\,m^{-1}$ for the monodromy
about $\a=\infty$ in the symplectic basis.
(For completeness we also record the result for $k=5$ already
given in \cdgp.)

$$
{{\tenpoint{
       \mat{k=5:\,\,&m=\pmat{-{3\over 5}&-{1\over 5}&
            \phantom{-}{21\over 5}&\phantom{-}{8\over 5}\cr
            \phantom{-}0&\phantom{-}0&-1&\phantom{-}0\cr
            -1&\phantom{-}0&\phantom{-}8&\phantom{-}3\cr
            \phantom{-}0&\phantom{-}1&-1&\phantom{-}0\cr}
            &A=\pmat{-9&-3&\phantom{-}5&\phantom{-}3\cr
                          \phantom{-}0&\phantom{-}1&\phantom{-}0&-1\cr
                          -20&-5&\phantom{-}11&\phantom{-}5\cr
                          -15&\phantom{-}5&\phantom{-}8&-4\cr}
            &S=\pmat{\phantom{-}51&\phantom{-}90&-25&0\cr
                           \phantom{-}0&\phantom{-}1&\phantom{-}0&0\cr
                           \phantom{-}100&\phantom{-}175&-49&1\cr
                           -75&-125&\phantom{-}35&1\cr}\cr
\noalign{\medskip}
                k=6:\,\,&m=\pmat{-{1\over 3}&-{1\over 3}&
            \phantom{-}{1\over 3}&\phantom{-}{1\over 3}\cr
            \phantom{-}0&\phantom{-}0&-1&\phantom{-}0\cr
            -1&\phantom{-}0&\phantom{-}3&\phantom{-}2\cr
            \phantom{-}0&\phantom{-}1&-1&\phantom{-}0\cr}
            &A=\pmat{\phantom{-}1&-1&\phantom{-}0&\phantom{-}1\cr
                          \phantom{-}0&\phantom{-}1&\phantom{-}0&-1\cr
                          -3&-3&\phantom{-}1&\phantom{-}3\cr
                          -6&\phantom{-}4&\phantom{-}1&-3\cr}
            &S=\pmat{\phantom{-}1&\phantom{-}6&\phantom{-}0&0\cr
                           \phantom{-}0&\phantom{-}1&\phantom{-}0&0\cr
                           \phantom{-}18&\phantom{-}81&\phantom{-}1&0\cr
                           -27&-129&-6&1\cr}\cr
\noalign{\medskip}
                k=8:\,\,&m=\pmat{-{1\over 2}&-{1\over 2}&
            \phantom{-}{1\over 2}&\phantom{-}{1\over 2}\cr
            \phantom{-}0&\phantom{-}0&-1&\phantom{-}0\cr
            -1&\phantom{-}0&\phantom{-}3&\phantom{-}2\cr
            \phantom{-}0&\phantom{-}1&-1&\phantom{-}0\cr}
            &A=\pmat{\phantom{-}1&-1&\phantom{-}0&\phantom{-}1\cr
                          \phantom{-}0&\phantom{-}1&\phantom{-}0&-1\cr
                          -2&-2&\phantom{-}1&\phantom{-}2\cr
                          -4&\phantom{-}4&\phantom{-}1&-3\cr}
            &S=\pmat{\phantom{-}1&\phantom{-}8&\phantom{-}0&0\cr
                           \phantom{-}0&\phantom{-}1&\phantom{-}0&0\cr
                           \phantom{-}16&\phantom{-}88&\phantom{-}1&0\cr
                           -40&-200&-8&1\cr}\cr
\noalign{\medskip}
                k=10:&m=\pmat{0&1&\phantom{-}1&\phantom{-}1\cr
             0&0&-1&\phantom{-}0\cr
             1&0&\phantom{-}0&-1\cr 0&1&-1&\phantom{-}0\cr}
            &A=\pmat{1&0&1&\phantom{-}0\cr
                          0&1&0&-1\cr 0&1&1&-1\cr 1&3&1&-2\cr}
            &S=\pmat{\phantom{-}1&\phantom{-}55&-10&0\cr
                           \phantom{-}0&\phantom{-}1&\phantom{-}0&0\cr
                           \phantom{-}0&-10&\phantom{-}1&0\cr
                           -10&-195&\phantom{-}45&1\cr}\cr}}}}
$$
The matrices $A$ satisfy $A^4=-1$ and $A^5=-1$ in the cases
$k=8$ and $k=10$, respectively.

The matrix $T=m\, t\, m^{-1}$, which describes the monodromy around $\a=1$,
is the same in all cases:
$$
{\tenpoint T=\pmatrix{1&0&0&0\cr 0&1&0&1\cr
                      0&0&1&0\cr 0&0&0&1\cr}}
$$
We have verified that the components of $\vec\Pi^\prime$ pass
the consistency check of ref.\cdgp,
$W_1=0$ where $W_k=z^a\partial_\a^k{\cal G}_a
-{\cal G}_a\partial_\a^k z^a$.

Let us now turn to the quantities which are relevant for
the low energy effective Lagrangian of the string theory
compactified of the Calabi-Yau spaces under considerations.
The K\"ahler potential on the one-dimensional moduli space
for the complex structure modulus is given in terms of the
prepotential:
\eqn\kahler{e^{-K}=-i\left(z^a\partial_{z^{\bar a}}\bar{\cal G}
-\bar z^a\partial_{z^a}{\cal G}\right)\,=\,-i\vec\Pi^\dagger
\Sigma\vec\Pi\,.}
We give the results in the limits $\a\to\infty$
for which we use the solutions in the form \solsinfty\
and for $\a\to 0$ using \solzero\.\hfil\break
\noindent $\a\to\infty$:
\eqn\kahlerinfty{e^{-K}\simeq{(2\pi)^3\over{\rm Ord}\,G}
            \left({4k\over 3}\log^3|\g\a|+{2\over 3k^2}
            \left(k^3-\sum_{i=0}^4\nu_i^3\right)\zeta(3)\right)}
\eqn\metricinfty{g_{\a\bar\a}\simeq{3\over 4|\a|^2\log^2|\g\a|}
\left(1+{2\left({\displaystyle\sum_{i=0}^4}\left({\nu_i\over k}\right)^3
-1\right)\zeta(3)\over\log^3|\g\a|}\right).}
In terms of the variable $t\propto i\log(\g\a)$
the leading behaviour is $g_{t\bar t}\simeq-{3\over(t-\bar t)^2}$
which is the metric for the upper half plane with curvature
$R=-{4/ 3}$.
\hfil\break
\noindent $\a\to 0$\footnote{$^*$}{For $k=10$ the analytic expressions
are rather cumbersome so we decided to give only the numerical
values.}:
\eqn\kahlerzero{
\mat{&e^{-K}_{k=5}=\displaystyle{{(2\pi)^3\over 5^5}{\Gamma^5({1\over5})\over
                   \Gamma^5({4\over 5})}\,|\a|^2\,+\,
                   O(|\a|^{4})};\quad&
     &e^{-K}_{k=6}=\displaystyle{{2^{13/3}\pi^8\over 3^{11/2}
                   \Gamma^2({2\over 3})
                  \Gamma^8({5\over 6})}\,|\a|^2\,+\,O(|\a|^{4})},\cr
\noalign{\medskip}
     &e^{-K}_{k=8}=\displaystyle{{\pi^7\over 128}{\cot^2({\pi\over 8})
                   \over\Gamma^8({7\over 8})}\,|\a|^2
                   \,+\,O(|\a|^{6})},&
     &e^{-K}_{k=10}\simeq 104.61\,|\a|^2\,+\,O(|\a|^{6});\cr}}

\eqn\metriczero{
\mat{
         &{g}^{k=5}_{\a \bar \a}=
         \displaystyle{ 25 \left(\Gamma({4\over 5}) \Gamma({2\over 5})\over
         \Gamma^3({1\over 5}) \Gamma({3\over 5})\right)^5+O(|\a|^2)},\qquad
         &{g}^{k=6}_{\a \bar \a}=\displaystyle{
         {3 \Gamma^8({5\over 6})\over
         {{2^{{2\over 3}}}\,{{\pi }^2}} \Gamma^4({2\over 3})}\,+ O(|\a|^2)},
         \qquad \cr
\noalign{\medskip}
         &{g}^{k=8}_{\a \bar \a}=\displaystyle{
         {64 (3-2^{3/2})^2 \Gamma^8({7\over 8})\over
          \Gamma^8({5\over 8})}\,\, |\a|^2\, + O(|\a|^8)},\,\,\qquad
          &{g}^{k=10}_{\a \bar \a}\simeq
          0.170 \,\, |\a|^2\, +\,O(|\a|^6).\qquad }}

The Yukawa couplings are $\kappa_{\a\a\a}=\int_{M^\prime}\Omega\wedge
{\partial^3\Omega\over\partial\a^3}$.
Decomposing the holomorphic three form $\Omega$ as in Appendix B and
using that $\int_{M^\prime}\a_a\wedge\b^b=
\delta_a^b,~\int_{M^\prime}\a_a\wedge\a_b
=\int_{M^\prime}\b^a\wedge\b^b=0$
one finds $\kappa_{\a\a\a}=W_3$. This form for
$\kappa_{\a\a\a}$ is not the most convenient one and we will
use it only to fix the overall normalization by evaluating
$W_3$ in the limit $\a\to 0$. Another relation satisfied
by $\kappa_{\a\a\a}$ has been found in \cdgp\ . There it is
shown that the Yukawa coupling satisfies the first order differential
$\partial_\a\kappa_{\a\a\a}+{1\over 2}C_3\kappa_{\a\a\a}=0$ where
$C_3$ is the coefficient of the third derivative in the
Picard-Fuchs equation with the coefficient of the highest
derivative normalized to one.
Using the form of the Picard-Fuchs equation as given in
eq.\pereq\ we easily derive $\kappa_{\a\a\a}\propto
{\a^{k-5}\over 1-\a^k}$. Taking now into account that below
eq.\solinfty\ we have made a gauge transformation by
rescaling the holomorphic three form by a factor $q(\a)\propto\a$,
which affects the Yukawa coupling as $\kappa_{\a\a\a}\to
q^2(\a)\kappa_{\a\a\a}$ and fixing the overall normalization as
described above, we finally obtain
\eqn\yukawas{\kappa_{\a\a\a}={(2\pi i)^3\over {\rm Ord}\,G}
             {k\a^{k-3}\over 1-\a^k}}
The invariant Yukawa couplings are defined as
\eqn\invyukawa{{\cal Y}_{inv}=g_{\a\bar\a}^{-3/2} e^K
|\kappa_{\a\a\a}|}
They correspond to a canonically normalized kinetic energy
of the matter fields (hence the factor $g_{\a\bar\a}^{-3/2}$)
and are invariant under K\"ahler gauge
transformations (hence the factor $e^K$).
In the limits considered above we find for the
leading terms
of the Yukawa couplings of the one multiplet of
$27$ of $E_6$:
\hfil\break\noindent$\a\to \infty$:
\eqn\yukinfty{{\cal Y}_{inv}={2\over\sqrt{3}}\,\quad\forall\,k\,.}
\noindent$\a\to 0$:
\eqn\yukzero{
   \mat{
         &{\cal Y}^{k=5}_{inv}=
         \displaystyle{\left(\Gamma^3({3\over 5}) \Gamma({1\over 5})\over
         \Gamma^3({2\over 5}) \Gamma({4\over 5})\right)^{5\over 2}+
          O(|\a|^2)},\qquad
         &{\cal Y}^{k=6}_{inv}=\displaystyle{2^{4\over 3}}\,\,
         |\a|  \,+ O(|\a|^3),
         \qquad \cr
\noalign{\medskip}
         &{\cal Y}^{k=8}_{inv}=\displaystyle{
        {\Gamma^6({5\over 8}) \Gamma^2({1\over 8})\over
         \Gamma^6({3\over 8}) \Gamma^2({7\over 8})}+O(|\a|^2)} ,\,\,\,\,\,\,
          &{\cal Y}^{k=10}_{inv}=
          3.394 \,\, |\a|^2\, +\,O(|\a|^6)\,.}}

For $k=5,8$ the nonvanishing couplings coincide with the values
of the corresponding Gepner models, which can be calculated using
the relation\ref\disgre{J. Distler and B. Greene, \npb309(1988)295.}
between the operator product coefficients of the
minimal $(n=2)$ superconformal models and the known ones of the
$su(2)$ Wess-Zumino-Witten theories. In the the $k=6,10$ cases the
additional $U(1)$ selection rules at the Gepner point exclude the coupling,
which is allowed for generic values of the modulus.

\newsec{The mirror maps}
So far we have only considered the complex structure modulus and the
K\"ahler metric and Yukawa coupling on the mirror manifold $M^\prime$ of
the original manifold $M$.
Here all quantities could be obtained from the
solutions of the Picard-Fuchs equations. The couplings involving
the (1,1) sector of moduli space of the original manifold
cannot be obtained from the periods on that manifold, but rather
{}From the periods of the mirror manifold via the mirror
map briefly described in the introduction.

The (1,1) sector of the original manifold is also described by a
holomorphic function, denoted by
${\cal F}$. It is also of the form ${\cal F}={1\over 2}
\omega^a{\cal F}_a$, which is homogeneous of degree two
in the $\omega^a$ and thus ${\cal F}_a=\partial{\cal F}/\partial
\omega^a$.

The large radius limit of ${\cal F}$ is known; it takes
the simple form ${\cal F}_0=-{\kappa_0\over 6}
{(\omega^1)^3\over\omega^2}=-{\kappa_0\over 6}(\omega^2)^2 t^3
=(w^2)^2\tilde{\cal F}_0$
where we have introduced the inhomogeneous coordinate
$t={\omega^1\over\omega^2}$. $\kappa_0=-\partial_t^3\tilde{\cal F}_0$
is the infinite radius limit of the Yukawa coupling
and is given by an intersection number.
In general, if we define ${\cal F}=(w^2)^2\tilde{\cal F}$ the
Yukawa coupling $\kappa$ and the K\"ahler potential are given
in terms of $\tilde{\cal F}$ as follows:
\eqn\fktilde{\eqalign{\kappa_{ttt}&=-\partial_{\omega^1}^3 {\cal F}
\Bigl|_{\omega^2=1}
=-\partial_t^3\tilde{\cal F}\cr
K&=-\log\left((t-\bar t)(\partial_t\tilde{\cal F}
-\bar\partial_{\bar t}\bar{\tilde{\cal F}})
-2(\tilde{\cal F}-\bar{\tilde{\cal F}})\right)\cr}}
(Here $K$ differs from the one given in terms ${\cal F}$
(cf. below) by a K\"ahler transformation.)

Let us briefly recall how to compute the intersection numbers
\ref\yau{S.-T. Yau, {\it Compact three dimensional K\"ahler manifolds
with zero Ricci curvature}, in Geometry, Anomalies, Topology,
W. Bardeen and A. White, eds., World Scientific 1985.}.
The K\"ahlerform $\tilde J$ of a complete intersection Calabi-Yau $M$,
which is defined
by a polynom constraint of degree $deg(p)$ in a weighted projective
space $\IP$ is given as the pullback of the
K\"ahlerform $J$ of the latter one, by  $\tilde J=i^* J$, where
$i:M \hookrightarrow \IP$
is the inclusion map. The topological three point function
$\kappa_0(\tilde J,\tilde J,\tilde J)$ is then most easily
computed by lifting
the integration over $M$ to the ambient space $\IP(\nu_0,\ldots,\nu_n)$
$$
\kappa_0(\tilde J,\tilde J,\tilde J)=\int_M
\tilde J\wedge\tilde J\wedge\tilde J=\int_{\IP(\nu_0,\ldots,\nu_n)}
J\wedge J \wedge J\wedge \eta_M
$$
using the Poincare dual
$\eta_M=deg(p) J$ of $M$ in $\IP(\nu_0,\ldots,\nu_n)$.
Taking into account the correct normalisation for the K\"ahlerform of the
$\IP(\nu_0,\ldots,\nu_n)$, namely such that
$\int_{\IP(\nu_0,\ldots,\nu_n)} J^n=(\prod\nu_i)^{-1}$,
we simply get
$$
\kappa_0(\tilde J,\tilde J,\tilde J)={deg(p)\over\prod_{i=0}^n \nu_i}\,.
$$
That is, $\kappa_0=\{5,3,2,1\}$ for $k=\{5,6,8,10\}$.

To get the K\"ahler potential we use eq.\kahler~
with the replacement $(z^a,{\cal G}_a)\to(\omega^a,{\cal F}_a)$,
or $\vec\Pi^\prime\to\vec\Pi=\left({\cal F}_1,{\cal F}_2,
\omega^1,\omega^2\right)^{\rm T}$.
We find $(t=t_1+it_2)$
\eqn\kahlerbare{K_0=-\log\left({4\kappa\over 3}t_2^3\right)\,.}
{}From this K\"ahler potential we easily derive the large
radius limits of the metric $g^0_{t\bar t}={3\over 4t_2^2}$
and of the Ricci tensor $R^0_{t\bar t}=-{2\over 3}g^0_{t\bar t}$.
For the Ricci scalar one thus gets $R^0=-{4\over 3}$ and
for the invariant Yukawa coupling ${\cal Y}_0={2\over\sqrt{3}}$.
These same constant values were found as the large complex structure
limits for the (2,1) moduli space of $M^\prime$.

These infinite radius results now get modified by sigma model
loops and instanton contributions, the latter being
non-perturbative in the sigma model expansion parameter
$1/R^2\sim 1/t$, $R$ being a measure for the size of the manifold.
This means that the prepotential in general
has the form
\eqn\ftilde{\tilde{\cal F}=-{\kappa_0\over6}t^3
+{1\over 2}at^2+bt+c+O\Bigl(e^{-t}\Bigr)}
leading to
$$
\vec\Pi=(\omega^2)^2\,\pmatrix{-{\kappa_0\over 2}t^2+at+b\cr
{\kappa_0\over 6}t^3+bt+c\cr t\cr 1\cr}\,.
$$
The polynomial part is perturbative. It is fixed by the fact
that for the Yukawa couplings there is a perturbative
non-renormalization theorem and only imaginary parts of
$a,b$ and $c$ do affect the K\"ahler metric.

The instanton corrections are in general hopeless to compute
directly. It is however made possible by the mirror
hypothesis which implies that the two prepotentials
${\cal G}$ and ${\cal F}$ are essentially the same,
but generally expressed
in two different symplectic bases; i.e. $\vec\Pi$ and $\vec\Pi^\prime$
are related by a symplectic transformation and then
lead to the same K\"ahler metric.
We find an integer symplectic matrix which relates $\vec\Pi$ and
$\vec\Pi^\prime$ up to a gauge transformation by relating their
asymptotic limits
where the limit of $\vec\Pi^\prime$ is obtained from the asymptotic
limit of \solsinfty\ ~and that of $\vec\Pi$ from the Ansatz
\ftilde.
We have already seen that in terms of the variable $t\propto
i\log(\g\a)$ the large complex structure and large
radius limits of the K\"ahler metrics for the moduli spaces of the
(2,1) and (1,1) moduli agree. Fixing the asymptotic relation to
$t\simeq\-{k\over 2\pi i}\log(\g\a)$ we find
for each case an integer symplectic matrix $N$ such that
${1\over\omega^2}\vec\Pi=N
{1\over{\cal G}_2}\vec\Pi^\prime$. This then also gives the
relation between
$\omega^1,\,\omega^2$ and
$w_0,\,w_1,\,w_2,\,w_{k-1}$
and allows us to express $t$ in terms of $\a$.
Choosing $N$ as simple as possible, we find
$$
N_{k=5}={\tenpoint{
        \pmatrix{-1&0&\phantom{-}0&\phantom{-}0\cr
                 \phantom{-}0&0&\phantom{-}0&-1\cr
                 \phantom{-}2&0&-1&\phantom{-}0\cr
                 \phantom{-}0&1&\phantom{-}0&\phantom{-}0\cr}\,,
}}
\quad
N_{k=6}\,=\,N_{k=8}=-\Sigma\,,
\quad
N_{k=10}={\tenpoint{
         \pmatrix{-1&0&\phantom{-}0&\phantom{-}0\cr
                  \phantom{-}0&0&\phantom{-}0&-1\cr
                  \phantom{-}0&0&-1&\phantom{-}0\cr
                  \phantom{-}0&1&\phantom{-}0&\phantom{-}0\cr}\,.
}}
$$
which corresponds to the following choice of the parameters
$a,b$ and $c$ in
\ftilde\footnote{$^*$}{The contribution $\propto\zeta(3)$
arises from the terms $\propto\phi^{\prime\prime}(0)$
in eq.\solsinfty.}
\eqn\abc{\{a,b,c\,\}=
\,\cases{\left\{-{11\over 2},{25\over 12},-{25i\over\pi^3}
\zeta(3)\right\}\qquad&$(k=5)$\cr
\noalign{\smallskip}
\left\{-{9\over2},{7\over 4},-{51i\over2\pi^3}\zeta(3)
\right\}&$(k=6)$\cr
\noalign{\smallskip}
\left\{-3,{11\over 6},-{37i\over\pi^3}\zeta(3)\right\}&$(k=8)$\cr
\noalign{\smallskip}
\left\{-{1\over2},{17\over 12},-{36i\over\pi^3}\zeta(3)\right\}
&$(k=10)$\cr}}
The relation between $t$ and $\a$ is
\def\o{\omega}
\eqn\mirrormap{t={\o^1\over\o^2}=-{k\over 2\pi i}\left\lbrace\log(\g\a)
+{{\displaystyle \sum_{N=0}^\infty}{(kN)!\over\prod_{i=0}^4(\nu_i N)!}
\phi(N)(\g\a)^{-kN}\over
{\displaystyle \sum_{N=0}^\infty}{(kN)!\over\prod_{i=0}^4(\nu_i N)!}
(\g\a)^{-kN}}\right\rbrace .}
where the second expression is valid for $\a$ large.
Using the known
transformation $(at)^{-1}$ on
$\vec w$
we have checked in all cases that $t(\a)$
transforms as $t\rightarrow t + 1$ and thus $s=(at)^{-k}:\,t\to t+k$
when $\a$ is transported around infinity.
This is in accordance with the Lemma in Section 2 of
ref. \morrison , where this fact is used to determine
$t$ up to an additive constant.

Note that besides the $O(t^3)$ term which fixes the large radius
limit of the Yukawa coupling among the polynomial terms
in $\tilde{\cal F}$ only the constant one contributes to the
K\"ahler metric. It has been identified in \cdgp\  with the
four loop contribution calculated in \ref\gvz{M. Grisaru,
A. van de Ven and D. Zanon, \plb173(1986)423, \npb277(1986)388,
\npb277(1986)409.}. This term also makes its appearance in
the effective low-energy string actions extracted from tree level
string scattering amplitudes\ref\grsl{D. Gross and J. Sloan,
\npb291(1987)41; N. Cai and C. Nunez, \npb287(1987)279.}.
The exponentially small instanton corrections do affect the
Yukawa couplings as well as the K\"ahler metric.
To get the Yukawa coupling we transform $\kappa_{\a\a\a}$
to the coordinate $t$ and find that the infinite radius
value $\kappa_0$ gets corrected to
\eqn\yukawat{\kappa_{ttt}=\left({\omega^2\over{\cal G}_2}\right)^2
\kappa_{\a\a\a}\left({d\a\over dt}\right)^3\,.}
The prefactor expresses the gauge freedom and is due to the
relative factor (besides the integer symplectic matrix)
between $\vec\Pi$ and $\vec\Pi^\prime$ whose components
appear in the definition of the holomorphic three form which
enters quadratically in $\kappa_{\a\a\a}$.
In the gauge $\omega^2=1$ this becomes $\kappa_0+O(q)$ with
$q=\exp(2\pi i t)$,
where the instanton contributions come with integer
coefficients.
Indeed, on inverting the series \mirrormap\
and expressing the result in the form
$\kappa_{ttt}=\kappa_0+\sum_{j=1}^\infty{n_j j^3 q^j\over 1-q^j}$
conjectured in \cdgp\  and proven in \ref\asmo{P.S. Aspinwall
and D.R. Morrison, {\it Topological field theory and
rational curves}, preprint OUTP-91-32P.} we find the numbers
$n_j$ which count the rational curves of degree $j$ in $M$
\footnote{$^*$}{Our results agree with those of \cdgp \ for
the case $k=5$ and with those of \morrison\
for the cases $k=5,6,8$. Our results for $k=10$ differ by a
factor of two, due to the different
normalizations of the constant term.}.
\break
$$\vbox{
{\tenpoint{
{\offinterlineskip\tabskip=0pt
\halign{\strut\vrule#&$#$~&
\vrule$#$&~\hfil$#$~&\vrule$#$&~\hfil$#$~&\vrule$#$\cr
\noalign{{\hrule}}
& &&\hfil k=5\quad \hfil && \hfil k=6\qquad\hfil &\cr
\noalign{\hrule}
&~n_0&&5&&3&\cr
&~n_1 &&2875
&& 7884
&\cr
&~n_2 && 609250
&& 6028452
&\cr
&~n_3 && 317206375
&& 11900417220
&\cr
&~n_4 && 242467530000
&& 34600752005688
&\cr
&~n_5 && 229305888887625
&& 124595034333130080
&\cr
&~n_6 && 248249742118022000
&& 513797193321737210316
&\cr
&~n_7 && 95091050570845659250
&& 2326721904320912944749252
&\cr
&~n_8 && 375632160937476603550000
&& 11284058913384803271372834984
&\cr
&~n_9 && 503840510416985243645106250
&& 57666069759834844985369823018768
&\cr
\noalign{{\hrule}}
& &&\hfil k=8 \quad\hfil && \hfil k=10 \qquad\hfil &\cr
\noalign{\hrule}
&~n_0&&2&&1&\cr
&~n_1
&& 29504
&& 231200
&\cr
&~n_2
&& 128834912
&& 12215785600
&\cr
&~n_3
&& 1423720546880
&& 1700894366474400
&\cr
&~n_4
&& 23193056024793312
&& 350154658851324656000
&\cr
&~n_5
&& 467876474625249316800
&& 89338191421813572850115680
&\cr
&~n_6
&& 10807872280363954752338400
&& 26107067114407746641915631734400
&\cr
&~n_7
&& 274144987164929172592851362112
&& 8377961119575977127785199800102445600
&\cr
&~n_8
&& 7446718087338043414223489290659040
&& 2879133858909474665080674534026982622960000
&\cr
&~n_9
&& 213140047760089302995646535567239779840
&& 1042529487474393188294680546419175097976102240000
&\cr
\noalign{\hrule}}}}}
}$$
\vfill\break

\newsec{Conclusions}
Starting from the Picard-Fuchs equations we have computed the
complete expressions for the prepotential, the K\"ahler
metric and Yukawa couplings of the heterotic string
compactified on those Calabi-Yau spaces with one (2,1)
modulus which can be represented by one polynomial constraint
in $\IP_4$. We could also get the same
quantities for the manifold related by mirror symmetry, where
the dependence is now on the (1,1) modulus and is due to
instanton corrections.

One can now also study, as was done for the $k=5$ case in
\cdgp\ , the duality symmetry of those models. One can easily
generalize the procedure of ref.\cdgp \ and identify a
parameter $\gamma\in H^+$, on which the monodromy transformations
on the periods are represented as a two-generator
discrete subgroup of $SL(2;\R)$.
$\gamma$ is a map of the $\a^k$ plane to a pair of triangles
which together constitute a fundamental region for the
group of duality transformations.
Since the two transformations $S$ and $T$ are of infinite order they
will correspond to two parabolic elements whereas the transformation
$A$ will be represented\footnote{$^{**}$}{ We use the
same letters to represent the
transformations on $\vec\Pi^\prime$ and on $\gamma$.} by an elliptic
element\footnote{$^{*}$}{A duality group
with two parabolic and one elliptic  element of order $3$ was
observed for the one dimensional $Z_3$ orbifold with a discrete Wilson
line\ref\erler{J. Erler, {\it Untersuchungen von
Modulir\"aumen in Stringtheorien}, Ph. D. Thesis TU-M\"unchen,
preprint MPI-Ph/92-21}.} of order $k$
\ref\lehner{A good reference on this subject is e.g.
J. Lehner, {\it A short course in automorphic functions},
Holt, Rinehart and Winston, 1966.}.
Since there are only two generators these elements will not be
independent. The angles of the two triangles will thus be
$(0,0,{\pi\over k})$ corresponding to fixed points of two
parabolic and one elliptic transformation.
We choose these fixed points to be
$i\infty$, $\tan(\pi\delta)$ and $i$.
Here we have defined $\delta={k-1\over 2k}$.
The first two fixed points belong to two parabolic
and the last to an elliptic element.
The function $\g(\a)$ then maps the
fixed points of $A$, $S$ and $T$ to
$\g(0)=i$, $\g(\infty)=i \infty$ and $\g(1)=\tan(\pi\d)$,
and can be determinded by standard methods
\ref\kleinfricke{See, e.g. F. Klein und R. Fricke,
{\it Vorlesungen \"uber die Theorie der elliptischen Modulfunktionen,
Vol. I}, Teubner 1890.}
to be a ratio of hypergeometric functions
$$
\eqalign{
\g&=i{Z_1-e^{2\pi i\d} Z_2\over Z_1+e^{2\pi i\d} Z_2}\cr
&={i\over\pi}\tan(\pi\d)\left\lbrace\log(\a^k)-i\pi+
{\displaystyle{\sum_{n=0}^\infty
{\Gamma(n\!+\!\d)\Gamma(n\!+\!1\!-\!\d)\over(n!)^2\a^{kn}}
\left(2\psi(1\!+\!n)-\psi(n\!+\!\d)
-\psi(1\!+\!n\!-\!\d)\right)}\over\displaystyle{
\sum_{n=0}^\infty{\Gamma(n+\d)\Gamma(n+1-\d)\over (n!)^2\a^{kn}}}}
\right\rbrace\cr}
$$
where the last expression is valid for $|\a|>1$. We have defined
$$
Z_1={\Gamma^2(\d)\over\Gamma(2\d)}
\, {\rm F}(\d,\d,2\d;\a^k)\,,\quad
Z_2={\Gamma^2(1\!-\!\d)\over
\Gamma(2(1\!-\!\d))}\a
\, {\rm F}(1\!-\!\d,1\!-\!\d,2(1\!-\!\d);\a^k).
$$
Using the analytic continuation properties of the hypergeometric
functions we find
$$
A=\pmatrix{\phantom{-}\cos(2\pi\d)&
-\sin(2\pi\d)\cr
\phantom{-}\sin(2\pi\d)&
\phantom{-}\cos(2\pi\d)\cr}\,,\quad
T=\pmatrix{\cos(2\pi\d)&2\tan(\pi\d)\sin^2(\pi\d)\cr
-\sin(2\pi\d)&1+2\sin^2(\pi\d)\cr}\,,
$$
with fixed points $i$ and $\tan(\pi\d)$, respectively
and
$$
AT=\pmatrix{1 & -2\tan(\pi\d)\cr
0&1\cr}\,.
$$
with fixed point $i\infty$.
The fundamental region $\Gamma$
can  be chosen to be the region of the upper half plane
bounded by the two lines $|{\rm Re}(\g)|=\tan(\pi\d)$ and two circular arcs
extending from $\g=i$ to $\g=\pm\tan(\pi\d)$ intersecting the
imaginary axis at an angle of $\pi\over k$ and the two boundary lines
at zero angle. In figure 1 we show on the lefthandside these fundamental
regions in the $\g$ plane together with its images under the
operation of $A,\ldots,A^k$.
Here and below we restrict ourselves to the cases $k=6,8,10$.
For $k=5$ we refer to \cdgp.

It is possible to relate the parameters $\gamma$
and $t$ as their expansions in powers of $\a$ are known,
and thus express the couplings on the original
manifold $M$ in terms of the parameter on which the
duality group acts in a simple way. Using this power series expansions
we can map the fundametal region from the $\g$ plane into the $t$ plane. The
images are quadrangles whose upper corners are at $i\infty$, the
lower corners corresponding to the Gepner points are at
${1\over2}(-1+i\sqrt{3}),{1\over2}(-1+i(1+\sqrt{2}))$ and
${1\over2}(-1+i\sqrt{5+2 \sqrt{5}})$, respectively.
These values follow easily from the relation between $t$ and the
$w_i$ and their expansion for $|\a|<1$.
The real parts of the
location of the right and left corners are $-{1/2}\pm {1/2}$. On the
righthandside of figure 1 we have plotted these images of $\Gamma$ together
with the images of the boundaries of the maps of
$\Gamma$ in the $\g$ plane.
For the $k=8$ and $10$ case  pairs of lines in the $\g$ plane are
identified in the $t$ plane. As it is clear from the fact that these
patterns do not repeat under the shift  $t\to t+1$, there are points in
image of $\Gamma$ which can be identified by an $A^mTA^n$ operation in the
$t$ plane. For $k=6$ and $k=10$ the images of the different
fundamental regions even overlap.

While it is the $\g$ on which the duality acts simply by fractional linear
transformations the $t$ parameter has a simple geometrical interpretation
in that $te=B+iJ$ is the complexified
K\"ahler form (see e.g. \ref\cade{
P. Candelas and X. de la Ossa, \npb355(1991)455.})
so that the volume
of the manifold scales like $Im(t)^3=:R^6$. From Fig. 1 it is clear that one
can restrict oneself to a part of the $t$ plane
which is bounded from below
by a minimal length $Im(t)=R^2$.

The Yukawa coupling will have a simple transformation
law under modular transformations. This follows from the fact
that the one matter superfield which is related to the
modulus $\g$ via world-sheet supersymmetry will transform
homogeneously and to have an invaraint  supergravity action
the Yukawa coupling must also transform
homogeneously\ref\flt{S. Ferrara,
D. L\"ust and S. Theisen, \plb233(1989)147.}.
Having computed the Yukawa couplings we thus have an explicit
function of the modulus, which, when raised to the
appropriate power, is also a candidate for a non-perturbative
superpotential for the modulus itself.
Of course, whereas for the modular group $SL(2;\Z)$ this function
is known to be more or less unique, practically nothing is
known about automorphic functions of the groups we encouter
here.
Work in this direction is in progress.

The models considered here represent only a very restricted class
and do not lead to any even remotely realistic low energy theory.
To make further progress one has to extend the analysis in
several directions. One is to consider models described by
higher dimensional projective varieties. There are a few
examples of this kind with $h_{(2,1)}=1$, which can be studied
as a first step in this direction. Another generalization is
to models defined by more than one polynomial constraint.
The other obvious direction
to go is to consider models with more than one modulus,
leading to partial differential equations for the periods.
This seems to be the hardest of the possible generalizations.

Another possiblity to arrive at the couplings of strings
on Calabi-Yau spaces might be to follow the method of
ref.\ref\ceva{S. Cecotti and C. Vafa, \npb367(1991)359.}.
In the cases we have considered here it leads to two coupled
non-linear differential equations satisfied by the K\"ahler
potential and the K\"ahler metric, which can be cast into the form
of the $SP(4)$ Toda equation whose solution is known.
\nobreak

We plan to come back to these issues in the future.
\vskip1cm
\noindent Acknowledgements: We thank M. G. Schmidt for usefull
discussions.
\vfil\break
\appendix{A}{}
$$
\vbox{
{\tenpoint{
{\offinterlineskip\tabskip=0pt
\halign{\strut\vrule#&\hfil$~#~$\hfil&
\vrule#&\hfil$~#~$\hfil&\vrule#&\hfil$~#~$\hfil&#&
\hfil$~#~$\hfil&\vrule#&\hfil$~#~$\hfil&#&
\hfil$~#~$\hfil&\vrule#&\hfil$~#~$&\vrule#\cr
\noalign{{\hrule}}
& \multispan{13} \hfil $M=\{x_i\in {\IP}(1,1,1,1,1)\,\,|\,\,
x_0^5+x_1^5+x_2^5+x_3^5+x_4^5=0\}$ \hfil &\cr
\noalign{\hrule}
& {\rm group} && {\rm generators} && \multispan3 $h^{(1,2)}$ &&
\multispan3 $h^{(1,1)}$ &&\hfil\chi\hfil&\cr
\noalign{\hrule}
& 1 && && 101 && && 1 && && -200 &\cr
\noalign{\hrule}
& \Z_5 && (1\,0\,0\,4\,0) &&
49 && (25) && 5 && (1) && -88 &\cr
\noalign{\hrule}
& \Z_5 && (1\,2\,3\,4\,0) &&
21 && (21) && 1 && (1) && -40 &\cr
\noalign{\hrule}
& \Z_5\times\Z_5 &&
(1\,0\,0\,4\,0)\times(1\,2\,3\,4\,0) &&
21 && (5) && 17 && (1) && -8 &\cr
\noalign{\hrule}
& \Z_5 && (1\,2\,2\,0\,0) &&
17 && (17) && 21 && (1) && 8&\cr
\noalign{\hrule}
& \Z_5\times\Z_5 &&
(1\,2\,3\,4\,0)\times(1\,0\,2\,2\,0) &&
1 && (1) && 21 && (1) && 40 &\cr
\noalign{\hrule}
& \Z_5\times\Z_5 &&
(1\,0\,0\,4\,0)\times(1\,0\,4\,0\,0) &&
5 && (5) && 49 && (1) && 88 &\cr
\noalign{\hrule}
& \Z_5\times\Z_5\times\Z_5 &&
(1\,0\,0\,4\,0)\times(1\,0\,4\,0\,0)
\times(1\,4\,0\,0\,0) &&
1 && (1) && 101 && (1) && 200 &\cr
\noalign{\hrule}}}}}
\medskip
{\tenpoint{
{\offinterlineskip\tabskip=0pt
\halign{\strut\vrule#&\hfil$~#~$\hfil&
\vrule#&\hfil$~#~$\hfil&\vrule#&\hfil$~#~$\hfil&#&
\hfil$~#~$\hfil&\vrule#&\hfil$~#~$\hfil&#&
\hfil$~#~$\hfil&\vrule#&\hfil$~#~$&\vrule#\cr
\noalign{\hrule}
& \multispan{13} \hfil $M=\{x_i\in {\IP}(2,1,1,1,1)\,\, |\,\,
x_0^3+x_1^6+x_2^6+x_3^6+x_4^6=0\}$ \hfil &\cr
\noalign{\hrule}
\noalign{\hrule}
& {\rm group} && {\rm generators} && \multispan3 $h^{(1,2)}$ &&
\multispan3 $h^{(1,1)}$ &&\hfil\chi\hfil&\cr
\noalign{\hrule}
& 1 &&  && 103 &&  && 1 && && -204 &\cr
\noalign{\hrule}
& \Z_2 && (0\,3\,3\,0\,0) &&
63 && (55) && 3 && (1) && -120 &\cr
\noalign{\hrule}
& \Z_3 && (0\,2\,2\,2\,0) &&
60 && (40) && 6 && (1) && -108 &\cr
\noalign{\hrule}
& \Z_3 && (0\,4\,2\,0\,0) &&
45 && (37) && 3 && (1) && -84 &\cr
\noalign{\hrule}
& \Z_6 && (0\,1\,0\,0\,5) &&
43 && (21) && 7 && (1) && -72 &\cr
\noalign{\hrule}
& \Z_2\times\Z_2 &&
(0\,3\,3\,0\,0)\times(0\,3\,0\,3\,0) &&
37 && (31) && 7 && (1) && -60 &\cr
\noalign{\hrule}
& \Z_3 && (1\,2\,2\,0\,0) &&
31 && (31) && 13 && (1) && -36 &\cr
\noalign{\hrule}
& \Z_3\times\Z_3 &&
(0\,2\,2\,2\,0)\times(0\,4\,2\,0\,0) &&
30 && (16) && 12 && (1) && -36 &\cr
\noalign{\hrule}
& \Z_6 && (0\,4\,1\,1\,0) &&
29 && (20) && 17 && (1) && -24 &\cr
\noalign{\hrule}
& \Z_6\times\Z_3 &&
(0\,1\,0\,0\,5)\times(1\,2\,2\,0\,0) &&
27 && (7) && 15 && (1) && -24 &\cr
\noalign{\hrule}
& \Z_6 && (0\,3\,2\,1\,0) &&
23 && (19) && 11 && (1) && -24 &\cr
\noalign{\hrule}
& \Z_6 && (1\,5\,3\,2\,0) &&
21 && (17) && 9 && (1) && -24 &\cr
\noalign{\hrule}
& \Z_3\times\Z_3 &&
(0\,2\,2\,2\,0)\times(0\,2\,0\,2\,2) &&
23 && (15) && 17 && (1) && -12 &\cr
\noalign{\hrule}
& \Z_3\times\Z_3 &&
(0\,4\,0\,2\,0)\times(1\,2\,2\,0\,0) &&
19 && (11) && 13 && (1) && -12 &\cr
\noalign{\hrule}
& \Z_2\times\Z_6 &&
(0\,3\,0\,3\,0)\times(2\,1\,1\,0\,0) &&
13 && (9) && 19 && (1) && 12 &\cr
\noalign{\hrule}
& \Z_6\times\Z_2 &&
(0\,1\,0\,0\,5)\times(0\,3\,3\,0\,0) &&
17 && (11) && 23 && (1) && 12 &\cr
\noalign{\hrule}
& \Z_6\times\Z_3 &&
(0\,3\,2\,0\,1)\times(0\,4\,0\,2\,0) &&
9 && (5) && 21 && (1) && 24 &\cr
\noalign{\hrule}
& \Z_3\times\Z_6 &&
(0\,2\,0\,2\,2)\times(0\,3\,2\,1\,0) &&
11 && (7) && 23 && (1) && 24 &\cr
\noalign{\hrule}
& \Z_6 && (2\,1\,1\,0\,0) &&
15 && (15) && 27 && (1) && 24 &\cr
\noalign{\hrule}
& \Z_3\times\Z_6 &&
(0\,2\,0\,2\,2)\times(0\,1\,0\,0\,5) &&
17 && (8) && 29 && (1) && 24 &\cr
\noalign{\hrule}
& \Z_2\times\Z_6 &&
(0\,3\,3\,0\,0)\times(0\,4\,1\,1\,0) &&
12 && (10) && 30 && (1) && 36 &\cr
\noalign{\hrule}
& \Z_6\times\Z_6 &&
(0\,1\,0\,0\,5)\times(0\,3\,2\,1\,0) &&
13 && (3) && 31 && (1) && 36 &\cr
\noalign{\hrule}
& \Z_3\times\Z_3\times\Z_3 &&
(0\,2\,2\,2\,0)\times(0\,2\,0\,2\,2)
\times(0\,2\,2\,0\,2) &&
7 && (7) && 37 && (1) && 60 &\cr
\noalign{\hrule}
& \Z_3\times \Z_6 &&
(0\,2\,2\,2\,0)\times(0\,1\,0\,0\,5) &&
7 && (7) && 43 && (1) && 72 &\cr
\noalign{\hrule}
& \Z_6\times\Z_6  &&
(0\,4\,1\,1\,0)\times(0\,4\,1\,0\,1) &&
3 && (3) && 45 && (1) && 84 &\cr
\noalign{\hrule}
& \Z_6\times\Z_6 &&
(0\,1\,0\,0\,5)\times(0\,1\,0\,5\,0) &&
6 && (4) && 60 && (1) && 108 &\cr
\noalign{\hrule}
&  \Z_3\times\Z_3\times\Z_6 &&
(0\,2\,2\,2\,0)\times(0\,2\,0\,2\,2)
\times(0\,1\,0\,0\,5) &&
3 && (3) && 63 && (1) && 120 &\cr
\noalign{\hrule}
& \Z_3\times\Z_6\times\Z_6 &&
(0\,2\,2\,2\,0)\times(0\,1\,0\,0\,5)
\times(0\,1\,0\,5\,0) &&
1 && (1) && 103 && (1) && 204 &\cr
\noalign{\hrule}}}}}}
$$

\vskip1cm

$$
\vbox{
{\tenpoint{
{\offinterlineskip\tabskip=0pt
\halign{\strut\vrule#&\hfil$~#~$\hfil&
\vrule#&\hfil$~#~$\hfil&\vrule#&\hfil$~#~$\hfil&#&
\hfil$~#~$\hfil&\vrule#&\hfil$~#~$\hfil&#&
\hfil$~#~$\hfil&\vrule#&\hfil$~#~$&\vrule#\cr
\noalign{{\hrule}}
& \multispan{13} \hfil $M=\{x_i\in {\IP}(4,1,1,1,1)\,\,|\,\,
x_0^2+x_1^8+x_2^8+x_3^8+x_4^8=0\}$ \hfil &\cr
\noalign{\hrule}
& {\rm order} && {\rm generators} && \multispan3 $h^{(1,2)}$ &&
\multispan3 $h^{(1,2)}$ &&\hfil\chi\hfil&\cr
\noalign{\hrule}
\noalign{\hrule}
& 1 && && 149 && && 1 && && -296 &\cr
\noalign{\hrule}
& \Z_8 && (0\,1\,5\,1\,1) &&
106 && (85) && 2 && (1) && -208 &\cr
\noalign{\hrule}
& \Z_2 && (0\,4\,4\,0\,0) &&
83 && (77) && 3 && (1) && -160 &\cr
\noalign{\hrule}
& \Z_2\times\Z_8 &&
(0\,4\,4\,0\,0)\times(0\,1\,5\,1\,1) &&
69 && (49) && 5 && (1) && -128 &\cr
\noalign{\hrule}
& \Z_4 && (0\,4\,2\,2\,0) &&
49 && (39) && 9 && (1) && -80 &\cr
\noalign{\hrule}
& \Z_2\times\Z_2 &&
(0\,4\,4\,0\,0)\times(0\,4\,0\,4\,0) &&
47 && (41) && 7 && (1) && -80 &\cr
\noalign{\hrule}
& \Z_8 && (0\,7\,5\,3\,1) &&
43 && (39) && 3 && (1) && -80 &\cr
\noalign{\hrule}
& \Z_8 && (0\,1\,0\,0\,7) &&
43 && (21) && 11 && (1) && -64 &\cr
\noalign{\hrule}
& \Z_4\times\Z_8 &&
(0\,4\,2\,2\,0)\times(0\,1\,5\,1\,1) &&
41 && (25) && 9 && (1) && -64 &\cr
\noalign{\hrule}
& \Z_8 && (0\,3\,3\,1\,1) &&
35 && (35) && 19 && (1) && -32 &\cr
\noalign{\hrule}
& \Z_8\times\Z_8 &&
(0\,1\,0\,0\,7)\times(0\,1\,5\,1\,1) &&
33 && (13) && 17 && (1) && -32 &\cr
\noalign{\hrule}
& \Z_4\times\Z_8 &&
(0\,4\,0\,2\,2)\times(0\,3\,3\,1\,1) &&
29 && (23) && 21 && (1) && -16 &\cr
\noalign{\hrule}
& \Z_4\times\Z_2 &&
(0\,4\,2\,2\,0)\times(0\,4\,4\,0\,0) &&
27 && (21) && 19 && (1) && -16 &\cr
\noalign{\hrule}
& \Z_8 && (0\,5\,2\,1\,0) &&
25 && (19) && 17 && (1) && -16 &\cr
\noalign{\hrule}
& \Z_8 && (0\,4\,3\,1\,0) &&
23 && (19) && 15 && (1) && -16 &\cr
\noalign{\hrule}
& \Z_8\times\Z_8 &&
(0\,2\,3\,3\,0)\times(0\,1\,5\,1\,1) &&
17 && (11) && 25 && (1) && 16 &\cr
\noalign{\hrule}
& \Z_8\times\Z_8 &&
(0\,4\,3\,0\,1)\times(0\,5\,2\,1\,0) &&
15 && (11) && 23 && (1) && 16 &\cr
\noalign{\hrule}
& \Z_4\times\Z_2\times\Z_8 &&
(0\,4\,0\,2\,2)\times
(0\,4\,0\,4\,0)\times
(0\,3\,3\,1\,1) &&
19 && (15) && 27 && (1) && 16 &\cr
\noalign{\hrule}
& \Z_4\times\Z_8 &&
(0\,4\,2\,2\,0)\times(0\,4\,3\,1\,0) &&
21 && (11) && 29 && (1) && 16 &\cr
\noalign{\hrule}
& \Z_8 && (0\,2\,3\,3\,0) &&
17 && (17) && 33 && (1) && 32 &\cr
\noalign{\hrule}
& \Z_8\times\Z_8 &&
(0\,1\,0\,0\,7)\times(0\,4\,3\,1\,0) &&
19 && (5) && 35 && (1) && 32 &\cr
\noalign{\hrule}
& \Z_8\times\Z_2 &&
(0\,2\,3\,3\,0)\times(0\,4\,4\,0\,0) &&
9 && (9) && 41 && (1) && 64 &\cr
\noalign{\hrule}
& \Z_8\times\Z_2 &&
(0\,1\,0\,0\,7)\times(0\,2\,3\,3\,0) &&
11 && (11) && 43 && (1) && 64 &\cr
\noalign{\hrule}
& \Z_8\times\Z_8 &&
(0\,2\,3\,0\,3)\times(0\,5\,2\,1\,0) &&
3 && (3) && 43 && (1) && 80 &\cr
\noalign{\hrule}
& \Z_4\times\Z_4\times\Z_8 &&
(0\,4\,2\,2\,0)\times
(0\,4\,0\,2\,2)\times
(0\,3\,3\,1\,1) &&
7 && (7) && 47 && (1) && 80 &\cr
\noalign{\hrule}
& \Z_8\times\Z_8\times\Z_2 &&
(0\,2\,3\,0\,3)\times(0\,1\,0\,7\,0)
\times(4\,4\,0\,0) &&
9 && (7) && 49 && (1) && 80 &\cr
\noalign{\hrule}
& \Z_8\times\Z_4 &&
(0\,1\,0\,7\,0)\times(0\,4\,2\,2\,0) &&
5 && (5) && 69 && (1) && 128 &\cr
\noalign{\hrule}
& \Z_8\times\Z_8\times\Z_4 &&
(0\,1\,0\,7\,0)\times(0\,2\,3\,0\,3)
\times(0\,4\,2\,2\,0) &&
3 && (3) && 83 && (1) && 160 &\cr
\noalign{\hrule}
& \Z_8\times\Z_8 &&
(0\,1\,0\,0\,7)\times(0\,1\,0\,7\,0) &&
2 && (2) && 106 && (1) && 208 &\cr
\noalign{\hrule}
& \Z_8\times\Z_8\times\Z_8 &&
(0\,1\,0\,0\,7)\times(0\,1\,0\,7\,0)
\times(0\,1\,7\,0\,0) &&
1 && (1) && 149 && (1) && 296 &\cr
\noalign{\hrule}}}}}}
$$
\vfil\break

$$
\vbox{
{\tenpoint{
{\offinterlineskip\tabskip=0pt
\halign{\strut\vrule#&\hfil$~#~$\hfil&
\vrule#&\hfil$~#~$\hfil&\vrule#&\hfil$~#~$\hfil&#&
\hfil$~#~$\hfil&\vrule#&\hfil$~#~$\hfil&#&
\hfil$~#~$\hfil&\vrule#&\hfil$~#~$&\vrule#\cr
\noalign{{\hrule}}
& \multispan{13}\hfil$M=\{x_i\in {\IP}(5,2,1,1,1)\,\,|\,\,
x_0^2+x_1^5+x_2^{10}+x_3^{10}+x_4^{10}=0\}$\hfil&\cr
\noalign{\hrule}
& {\rm order} && {\rm generators} && \multispan3 $h^{(1,2)}$ &&
\multispan3 $h^{(1,1)}$ &&\hfil\chi\hfil&\cr
\noalign{\hrule}
\noalign{\hrule}
& 1 && && 145 && && 1 && && -288 &\cr
\noalign{\hrule}
& \Z_2 && (0\,0\,5\,5\,0) &&
99 && (81) && 3 && (1) && -192 &\cr
\noalign{\hrule}
& \Z_2\times\Z_2 &&
(0\,0\,5\,5\,0)\times(0\,0\,0\,5\,5) &&
67 && (49) && 7 && (1) && -120 &\cr
\noalign{\hrule}
& \Z_5 && (0\,1\,4\,4\,0) &&
47 && (31) && 11 && (1) && -72 &\cr
\noalign{\hrule}
& \Z_5 && (0\,0\,2\,8\,0) &&
37 && (29) && 13 && (1) && -48 &\cr
\noalign{\hrule}
& \Z_{10} && (0\,0\,1\,0\,9)  &&
39 && (17) && 15 && (1) && -48 &\cr
\noalign{\hrule}
& \Z_{10} && (0\,1\,3\,5\,0) &&
29 && (17) && 17 && (1) && -24 &\cr
\noalign{\hrule}
& \Z_{10} && (0\,1\,1\,7\,0) &&
17 && (15) && 29 && (1) && 24 &\cr
\noalign{\hrule}
& \Z_{10} && (0\,2\,3\,3\,0) &&
15 && (15) && 39 && (1) && 48 &\cr
\noalign{\hrule}
& \Z_2\times\Z_{10} &&
(0\,0\,0\,5\,5)\times(0\,1\,3\,5\,0) &&
13 && (9) && 37 && (1) && 48 &\cr
\noalign{\hrule}
& \Z_{10}\times\Z_2 &&
(0\,0\,1\,0\,9)\times(0\,0\,5\,5\,0) &&
11 && (9) && 47 && (1) && 72 &\cr
\noalign{\hrule}
& \Z_5\times\Z_5 &&
(0\,0\,2\,8\,0)\times(0\,0\,2\,0\,8) &&
7 && (7) && 67 && (1) && 120 &\cr
\noalign{\hrule}
& \Z_{10}\times\Z_5 &&
(0\,0\,1\,0\,9)\times(0\,0\,2\,8\,0) &&
3 && (3) && 99 && (1) && 192 &\cr
\noalign{\hrule}
& \Z_{10}\times\Z_{10} &&
(0\,0\,1\,0\,9)\times(0\,0\,1\,9\,0) &&
1 && (1) && 145 && (1) && 288 &\cr
\noalign{\hrule}}}}}}
$$
\vskip1cm

\appendix{B}{}
We follow ref.\cdgp\
and compute two of the periods explicitly
in a symplectic basis thus identifying
them with two particular solutions of the period equations.
The remaining two are then determined by requiring that the
monodromy transformations are represented as $Sp(4;\Z)$.
To do this we represent the periods as integrals of the holomorphic
three form over the cycles of the Calabi-Yau manifold.
We take $\Omega$ in the gauge\footnote{$^*$}{For the relation of
above form of the holomorphic three form with the one
implicit in our discussion at the beginning of section 3, we
refer to \ref\candelas{P. Candelas, \npb298(1988)458.}.}
\eqn\omega{\Omega(\a)=k\a{x_4 dx_0\wedge dx_1\wedge dx_2\over
\partial W/\partial x_3}\,.}
$\Omega$ can be expanded in a cohomology basis as
$\Omega=z^a\a_a-{\cal G}_a\b^a$ where the forms $\a_a$ and $\b^a$
are dual to the integer symplectic homology basis
$A^a$ and $B_a$ whic satisfies $A^a\bigcap B_b=\delta^a_b,\,
A^a\bigcap A^b=B_a\bigcap B_b=0$ and $\int_{A_a}\a_b=
\int_{B_a}\b^b=\delta_a^b$ with the other integrals vanishing.
With this the periods are $z^a=\int_{A^a}\Omega,\,{\cal G}_a
=\int_{B_a}\Omega$.
We now integrate the holomorphic three--form
$\Omega(\alpha)$ over the two cycles $A^2$ and $B_2$ which are
close to the node and which have intersection number one\cdgp\ .
They are defined as
\eqn\cycles{\eqalign{
   A^2&=\{x_k|x_4=1, x_i~{\rm real}, x_3~s.t.~W(x)=0\cong S^3~
       {\rm as}~\alpha\to 1\}\cr
   B_2&=\{x_k|x_4=1, |x_0|=|x_1|=|x_2|=\delta, x_3\to 0~{\rm as}~
       \alpha\to\infty\}\cr}}
To get the correct normalization for the periods we work in a
coordinate patch where one of the coordinates with weight
one is set to one. This completely fixes the equivalences
in $\IP_4$.

$z^2(\alpha)$: We will only need $z^2$ to lowest
order in $\epsilon^2:=(\alpha-1)$. For points close to the node
we set $x_i=1+y_i~(i=0,1,2,3)$
with $y_i=O(\epsilon)$ and define
$\partial W/\partial x_3=k(k-1)w_3+O(\epsilon^2)$ where $w_4$ is
linear in the $y_i$. One then finds that $W=0$ corresponds,
upon dropping terms $O(\epsilon^3)$, to
$(w_3^2+B_{}y_i y_j)=\left({2k\over k-1}\right)\epsilon^2$,
where $B_{ij}$ is a real symmetric matrix with positive eigenvalues.
Converting to spherical coordinates one then finds that
\eqn\zzwei{\eqalign{z^2(\alpha)=\int_{A^2}\Omega
   &={4k\pi^2\over(k-1)^2}(\det B) (\alpha-1)+\dots\cr
   &={4\pi^3\over k^{3/2}}\left(\prod_{i=0}^4\nu_i\right)^{1/2}
   (\alpha-1)\,+\,O\left((\a-1)^2\right)\cr}}
This shows that $z^2$ is the solution of the period equation
with index one at $\alpha=1$ which is free from logarithms.

$G_2(\alpha)$:
We define $\xi$ via $x_3=(\alpha x_0 x_1 x_2)^{1/(k-1)}\xi$
in terms of which $W=0$ reads $\xi=\epsilon+{\xi^k\over k}$
with $\e={1+W_0-x_3^k-x_4^k\over k(\alpha x_0 x_1 x_2)^{k/(k-1)}}$.
The holomorphic three-form then becomes
$\Omega={dx_0\wedge dx_1\wedge dx_2\over x_0 x_1 x_2(\xi^{k-1}-1)}$.
Expanding ${1\over 1-\xi^{k-1}}$ in powers of $\epsilon$ and
performing the integrals over $x_i$ with $|x_i|=\delta$ $(i=0,1,2)$
one finds\footnote{$^*$}{Here we use that given
$\xi=\epsilon+{\xi^k\over k}$, ${1\over 1-\xi^{k-1}}=
\sum_{m=0}^\infty{mk\choose m}{1\over k^m}\epsilon^{m(k-1)}$.
This is easily shown as follows: one makes the Ansatz
${1\over 1-\xi^{k-1}}=\sum_{\nu=0}^\infty a_\nu\epsilon^\nu$.
Noting that $1-\xi^{k-1}={\partial\over\partial\xi}\e(\xi)$.
and $a_\nu=\oint{d\epsilon\over 1-\xi^{\nu+1}}{1\over\e^{\nu+1}}
=\oint{d\xi\over\epsilon^{\nu+1}}$ we get the result
upon expanding ${1\over\epsilon^{\nu+1}}$ in powers of $\xi$ and
extracting the residue.}
\eqn\gzwei{{\cal G}_2={(2\pi i)^3\over{\rm ord}(G)}
\sum_{m=0}^\infty{(km)!\over\prod_{i=0}^4(m \nu_i)!}
(\gamma\alpha)^{-km}}
Candelas et al.\cdgp\ argue that under transport about $\a=1$
only the cycle $B_2$ transforms non-trivially in that it picks
up a multiple of $A^2$. This means that the period ${\cal G}_2$
will pick a multiple of the period $z^2$. This leads to the
identification of $\tilde w(\a)$ in eq.\soleins\ and the
assertion that the other periods are free from logarithms.

\vskip2cm
{\it Note added:}
While we were in the process of writing up these results we
were informed by D. L\"ust and P. Candelas that A. Font has also treated
the same one modulus models with similar methods\ref\font{A. Font, {\it
Periods and Duality Symmetries in Calabi-Yau Compactifications}, preprint
UCVFC-1-92.}.

\footatend\vfill\eject\immediate\closeout\rfile\writestoppt
\baselineskip=14pt\centerline{{\bf References}}\bigskip{\frenchspacing%
\parindent=20pt\escapechar=` \input refs.tmp\vfill\eject}\nonfrenchspacing

\end